\documentclass{elsart}%
\usepackage{amsmath}
\usepackage{graphicx}
\usepackage[centerlast,small]{caption2}
\usepackage{indentfirst}
\usepackage{amsfonts}
\usepackage{amssymb}%
\providecommand{\U}[1]{\protect\rule{.1in}{.1in}}
\DeclareMathOperator{\Li}{Li_{2}}
\begin{document}
\begin{frontmatter}
\title{Radiative corrections to the\\
Reggeized quark -- Reggeized quark -- gluon effective vertex\thanksref{RFBR}}
\thanks[RFBR]{Work supported by the Russian Fund for Basic Research,
project 04-02-16685-a.}
\author{A.V.~Bogdan}
\ead{A.V.Bogdan@inp.nsk.su} and \author{A.V.~Grabovsky}
\ead{A.V.Grabovsky@inp.nsk.su}
\address{Budker Institute of Nuclear Physics, 630090 Novosibirsk, Russia}
\begin{abstract}
This paper is devoted to the calculation of the Reggeized quark -- Reggeized quark -- gluon
effective vertex in perturbative QCD in the next--to--leading order. The case of QCD with massless
quarks is considered and the correction is obtained in the $D\rightarrow 4$ limit. This vertex
appears in the quark Reggeization theory, which next--to--leading order extension is now under
construction.
\end{abstract}
\end{frontmatter}

\section{Introduction}

The construction of the quark Reggeization theory in the next--to--leading
approximation (NLA) involves both the calculation of the next--to--leading
order (NLO) corrections to the quark effective vertices and trajectory and the
proof of the Reggeization hypothesis. Some of these tasks have already been
done. Firstly, all multiparticle Reggeon vertices required in the NLA were
obtained \cite{Lipatov:2000se}. It enabled one to prove the quark Reggeization
hypothesis in the quasi--multi--Regge kinematics (QMRK) \cite{Bogdan:2006wq},
important in the NLA. Next, NLO corrections to the effective
particle--particle--Reggeon (PPR) vertices, appearing in the leading
logarithmic approximation (LLA) were found \cite{FF01}. It allowed one to
obtain the NLO correction to the quark Regge trajectory \cite{BD-DFG}, which
was also a test of the hypothesis in the NLO but only for a particular elastic process.

In this paper the NLO correction to the coupling of two Reggeized quarks with
external gluon is calculated. The case of QCD with massless quarks is
considered and the correction is obtained in the ($D=4-2\epsilon)\rightarrow4$ limit.

The paper is organized as follows. In Section 2 we introduce the Reggeized
form of the gluon production amplitude required by analyticity, unitarity and
crossing symmetry and calculate the part of the effective vertex contributing
to the discontinuities of the production amplitude in $s_{1}$ and $s_{2}$
channels. In Section 3 we obtain the correction to the gluon production
amplitude in the framework of dispersive approach in $t_{1}$ channel. In
Section 4 we calculate one--loop corrections to quark-photon-Reggeized quark
vertex. In Section 5 we present the result for the Reggeized quark --
Reggeized quark -- gluon effective vertex.

We perform calculations for massless quarks, but for generality we give the
main formulae with $m$ kept.

\section{Production amplitude in the multi--Regge kinematics}

\begin{figure}[h]
\centering
\includegraphics{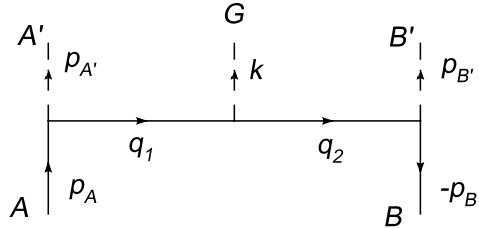}\caption{Schematic gluon production diagram.}%
\label{fig1}%
\end{figure}

Let us consider the high energy process $A+B\rightarrow A^{\prime}%
+G+B^{\prime}$ of a gluon $G$ production in the multi--Regge kinematics (MRK),
which means that all participating particles are well separated in the
rapidity space and have limited transverse momenta (see fig. 1):%
\begin{align}
(p_{A}+p_{B})^{2}  &  =s\gg s_{i}\gg|t_{j}|\,,\quad s_{1}=(p_{A^{\prime}%
}+k)^{2}\,,\quad s_{2}=(p_{B^{\prime}}+k)^{2}\,,\nonumber\\
t_{j}  &  =q_{j}^{2}\,,\quad q_{1}=p_{A}-p_{A^{\prime}}\,,\quad q_{2}%
=p_{B^{\prime}}-p_{B}\,,\nonumber\\
k  &  =q_{1}-q_{2}\,,\quad s_{1}s_{2}\simeq-sk_{\bot}^{2}\,. \label{MRK-1}%
\end{align}
Here $k_{\bot}$ is the gluon momentum component transverse to the plane
$p_{A},p_{B}$:%
\begin{equation}
k=\alpha p_{B}+\beta p_{A}+k_{\bot}\,,\quad\mathbf{k}^{2}=-k_{\bot}^{2}
\label{Sudakov}%
\end{equation}
and the last relation in eq.(\ref{MRK-1}) for the product $s_{1}s_{2}$ is a
consequence of the reality of the massless gluon:
\begin{equation}
k^{2}=s\alpha\beta+k_{\bot}^{2}=0\,.
\end{equation}
The behavior of the amplitude in the MRK is determined by the exchanges in
$q_{i}$ channels. For the case of quark quantum numbers in $q_{1}$ and $q_{2}$
channels, the multi--Regge form of the amplitude proved for the LLA in
\cite{Bogdan:2006af} is%
\begin{equation}
A_{2\rightarrow3}=\bar{\Gamma}_{B^{\prime}B}\frac{s_{2}^{\delta_{2}}}%
{m-\hat{q}_{2}}\,\gamma_{\text{Born}}^{G}\,\frac{s_{1}^{\delta_{1}}}{m-\hat
{q}_{1}}\Gamma_{A^{\prime}A}\,\,, \label{A-LLA}%
\end{equation}
where $\bar{\Gamma}_{P^{\prime}P}$ and $\Gamma_{P^{\prime}P}$ are the
particle-particle-Reggeon (PPR) effective vertices, describing $P\rightarrow
P^{\prime}$ transitions due to the interaction with Reggeons and taken in tree
approximation$;$ $\delta_{i}\equiv\delta(q_{i})$ is the quark Regge trajectory
\cite{FS}
\begin{equation}
\delta(q)=g^{2}C_{F}\int\frac{\mathtt{d}^{D-2}r_{\bot}}{(2\pi)^{D-1}}%
\frac{(m-\hat{q}_{\bot})}{(m-\hat{r}_{\bot})(q-r)_{\bot}^{2}}\,+\mathcal{O}%
(g^{4})\,. \label{trajectory}%
\end{equation}
Here, the space--time dimension $D=4-2\epsilon$, $C_{F}=N/2-1/(2N)$, with the
number of colors $N=3$ for QCD. $\gamma_{\text{Born}}$ =$\gamma_{\text{Born}%
}(q_{1},q_{2})$ is the Reggeon--Reggeon--particle (RRP) effective vertex,
describing the production of the gluon $G$ at Reggeized quark transitions
\cite{FS}
\begin{equation}
\gamma_{\text{Born}}^{G}=-gt^{G}e_{G}^{\ast\mu}\left(  \gamma^{\mu}+\left(
m-\hat{q}_{1}\right)  \frac{2\,p_{A}^{\mu}}{s_{1}}-\left(  m-\hat{q}%
_{2}\right)  \frac{2p_{B}^{\mu}}{s_{2}}\right)  \,, \label{gamma-Born}%
\end{equation}
with $t^{G}$ being the color group generator. Amplitude (\ref{A-LLA})
coincides with its real part because its imaginary part is subleading in the
LLA, but it is not quite so for amplitudes in the NLA.

If the Reggeization hypothesis in the NLA is correct then, assuming Regge
behavior in two subchannels $s_{1}$ and $s_{2}$ (see Fig.\ref{fig1}) with the
Reggeized quarks in the corresponding crossing channels $t_{1}$ and $t_{2}$,
from the general requirements of analyticity, unitarity and crossing symmetry
one obtains \cite{Bartels_1} the multi--Regge form
\begin{align}
&  A_{2\rightarrow3}=\frac{1}{4}\bar{\Gamma}_{B^{\prime}B}\frac{1}{m-\hat
{q}_{2}}\left\{  \left[  \left(  \frac{s}{\mathbf{-}t_{2}}\right)
^{\delta_{2}}+\left(  \frac{-s}{\mathbf{-}t_{2}}\right)  ^{\delta_{2}}\right]
\right. \nonumber\\
&  \times\left[  \left(  \frac{s_{1}}{\sqrt{(-t_{2})\mathbf{k}^{2}}}\right)
^{-\delta_{2}}\!\!\mathcal{R}\left(  \frac{s_{1}}{\sqrt{\mathbf{k}^{2}%
(-t_{1})}}\right)  ^{\delta_{1}}+\left(  \frac{-s_{1}}{\sqrt{(-t_{2}%
)\mathbf{k}^{2}}}\right)  ^{-\delta_{2}}\!\!\mathcal{R}\left(  \frac{-s_{1}%
}{\sqrt{\mathbf{k}^{2}(-t_{1})}}\right)  ^{\delta_{1}}\right] \nonumber\\
&  +\left[  \left(  \frac{s_{2}}{\sqrt{(-t_{2})\mathbf{k}^{2}}}\right)
^{\delta_{2}}\!\!\mathcal{L}\left(  \frac{s_{2}}{\sqrt{\mathbf{k}^{2}(-t_{1}%
)}}\right)  ^{-\delta_{1}}+\left(  \frac{-s_{2}}{\sqrt{(-t_{2})\mathbf{k}^{2}%
}}\right)  ^{\delta_{2}}\!\!\mathcal{L}\left(  \frac{-s_{2}}{\sqrt
{\mathbf{k}^{2}(-t_{1})}}\right)  ^{-\delta_{1}}\right] \label{A-NLA}\\
&  \times\left.  \left[  \left(  \frac{s}{-t_{1}}\right)  ^{\delta_{1}%
}+\left(  \frac{-s}{-t_{1}}\right)  ^{\delta_{1}}\right]  \right\}  \frac
{1}{m-\hat{q}_{1}}\Gamma_{A^{\prime}A}\,,\nonumber
\end{align}
where PPR vertices $\Gamma_{P^{\prime}P}=\Gamma_{P^{\prime}P}(t)$ depend on
the polarization of the particles $P$, $P^{\prime}$ and the squared momentum
transfer $t$. They are real for $t<0$. Hereafter, such expressions as
$(-s_{i})^{\delta}$ should be read as $(-s_{i}-\mathtt{i}o)^{\delta}$ so that
for $s_{i}>0$, $\ln(-s_{i})=\ln(s_{i})-\mathtt{i}\pi$. The RRP vertices
$\mathcal{R}=\mathcal{R}(q_{1},q_{2})$ and $\mathcal{L}=\mathcal{L}%
(q_{1},q_{2})$ depend on the polarization of the gluon and momenta $q_{1}$,
$q_{2}$. These vertices are real in all channels where $t_{i}<0$ and
$\mathbf{k}^{2}>0$. Moreover, in NLO both PPR and RRP vertices become
dependent on the energy normalization scale. Since the amplitude
$A_{2\rightarrow3}$ is physical, it does not depend on this scale. Therefore,
the energy normalization points in eq.(\ref{A-NLA}) may be chosen in an
arbitrary way supposing that the corresponding PPR and RRP vertices are
calculated at the same points. Our choice $-t_{j}$ and $\sqrt{\mathbf{k}%
^{2}(-t_{j})}\,$yields a particularly symmetric real part of the amplitude
$A_{2\rightarrow3}$ in the NLA:%
\begin{equation}
\operatorname{Re}A_{2\rightarrow3}=\bar{\Gamma}_{B^{\prime}B}\frac{1}%
{m-\hat{q}_{2}}\left(  \frac{s_{2}}{\sqrt{(-t_{2})\mathbf{k}^{2}}}\right)
^{\delta_{2}}(\mathcal{R}+\mathcal{L})\left(  \frac{s_{1}}{\sqrt
{\mathbf{k}^{2}(-t_{1})}}\right)  ^{\delta_{1}}\frac{1}{m-\hat{q}_{1}}%
\Gamma_{A^{\prime}A}\,.
\end{equation}
Note, that because of freedom in choosing the energy normalization, one has to
make sure that all PPR and RRP vertices in the amplitude are calculated at the
proper normalization points.

It is clear from eq.(\ref{A-NLA}) that the contribution of the sum
$\mathcal{R}+\mathcal{L}$ is leading, while that of the difference
$\mathcal{R}-\mathcal{L}$ is subleading. Indeed, in the LLA the imaginary part
of the amplitude is negligible, therefore, only the sum $\mathcal{R}%
+\mathcal{L}$ contributes to the effective vertex in Born approximation:%
\begin{equation}
\mathcal{R}^{(0)}+\mathcal{L}^{(0)}=\gamma_{\text{Born}}^{G}\,.
\end{equation}
On the contrary, if the Reggeization is valid, the difference $\mathcal{R}%
^{(0)}-\mathcal{L}^{(0)}$ at the same order $g$ contributes to the amplitude
(\ref{A-NLA}) only as a radiative correction. It can be obtained together with
the order $g^{3}$ corrections to the sum $\mathcal{R}^{(1)}+\mathcal{L}^{(1)}$
from the analysis of the NLO gluon production amplitude. We compare projection
of this amplitude on the color triplet state in $t_{i}$ channels taken with
the positive signature, with its multi--Regge form, assuming that the one loop
corrections $\Gamma_{P^{\prime}P}^{(1)}$ are known. In fact, with such
accuracy we get from eq.(\ref{A-NLA})%
\begin{align}
&  A_{2\rightarrow3}(\text{one-loop})=\bar{\Gamma}_{B^{\prime}B}^{(0)}\frac
{1}{m-\hat{q}_{2}}\gamma_{\text{Born}}^{G}\frac{1}{m-\hat{q}_{1}}%
\Gamma_{A^{\prime}A}^{(1)}+\bar{\Gamma}_{B^{\prime}B}^{(1)}\frac{1}{m-\hat
{q}_{2}}\gamma_{\text{Born}}^{G}\frac{1}{m-\hat{q}_{1}}\Gamma_{A^{\prime}%
A}^{(0)}\nonumber\\
&  +\frac{1}{4}\bar{\Gamma}_{B^{\prime}B}^{(0)}\frac{1}{m-\hat{q}_{2}%
}\nonumber\\
&  \times\left\{  \gamma_{\text{Born}}^{G}\delta_{1}\ln\left(  \frac
{s(-s)}{(-t_{1})^{2}}\right)  +\delta_{2}\gamma_{\text{Born}}^{G}\ln\left(
\frac{s(-s)}{(-t_{2})^{2}}\right)  +\left(  \gamma_{\text{Born}}^{G}\delta
_{1}-\delta_{2}\gamma_{\text{Born}}^{G}\right)  \ln\left(  \frac{s_{1}%
(-s_{1})}{s_{2}(-s_{2})}\right)  \right.  \label{A one-loop0}\\
&  +\left.  \left[  \left(  \mathcal{R}^{(0)}-\mathcal{L}^{(0)}\right)
\delta_{1}-\delta_{2}\left(  \mathcal{R}^{(0)}-\mathcal{L}^{(0)}\right)
\right]  \ln\left(  \frac{s_{1}(-s_{1})s_{2}(-s_{2})}{s(-s)(\mathbf{k}%
^{2})^{2}}\right)  +4\left(  \mathcal{R}^{(1)}+\mathcal{L}^{(1)}\right)
\right\}  \frac{1}{m-\hat{q}_{1}}\Gamma_{A^{\prime}A}^{(0)}\,.\nonumber
\end{align}
For massless QCD $\delta_{i}$ is \ a scalar and (\ref{A one-loop0}) simplifies
to%
\begin{align}
&  A_{2\rightarrow3}(\text{one-loop})-\bar{\Gamma}_{B^{\prime}B}^{(0)}\frac
{1}{\hat{q}_{2}}\gamma_{\text{Born}}^{G}\frac{1}{\hat{q}_{1}}\Gamma
_{A^{\prime}A}^{(1)}-\bar{\Gamma}_{B^{\prime}B}^{(1)}\frac{1}{\hat{q}_{2}%
}\gamma_{\text{Born}}^{G}\frac{1}{\hat{q}_{1}}\Gamma_{A^{\prime}A}%
^{(0)}\nonumber\\
&  -\frac{1}{4}\bar{\Gamma}_{B^{\prime}B}^{(0)}\frac{1}{\hat{q}_{2}}%
\gamma_{\text{Born}}^{G}\frac{1}{\hat{q}_{1}}\Gamma_{A^{\prime}A}^{(0)}\left[
\delta_{1}\ln\left(  \frac{s(-s)}{(-t_{1})^{2}}\right)  +\delta_{2}\ln\left(
\frac{s(-s)}{(-t_{2})^{2}}\right)  +(\delta_{1}-\delta_{2})\ln\left(
\frac{s_{1}(-s_{1})}{s_{2}(-s_{2})}\right)  \right]  \label{A one-loop}\\
&  =\bar{\Gamma}_{B^{\prime}B}^{(0)}\frac{1}{\hat{q}_{2}}\left[  \left(
\mathcal{R}^{(0)}-\mathcal{L}^{(0)}\right)  \frac{\delta_{1}-\delta_{2}}{4}%
\ln\left(  \frac{s_{1}(-s_{1})s_{2}(-s_{2})}{s(-s)(\mathbf{k}^{2})^{2}%
}\right)  +\mathcal{R}^{(1)}+\mathcal{L}^{(1)}\right]  \frac{1}{\hat{q}_{1}%
}\Gamma_{A^{\prime}A}^{(0)}\,.\nonumber
\end{align}

Eq.(\ref{A one-loop0}) states that the difference $\mathcal{R}^{(0)}%
-\mathcal{L}^{(0)}$ contributes to the discontinuities of $A_{2\rightarrow3}$
in $s_{1}$ and $s_{2}$ channels. For small $g$ we obtain%
\begin{equation}
(disc_{s_{1}}+disc_{s_{2}})A_{2\rightarrow3}=-\pi\mathtt{i}\,\bar{\Gamma
}_{B^{\prime}B}^{(0)}\frac{1}{m-\hat{q}_{2}}\left[  \left(  \mathcal{R}%
^{(0)}-\mathcal{L}^{(0)}\right)  \delta_{1}-\delta_{2}\left(  \mathcal{R}%
^{(0)}-\mathcal{L}^{(0)}\right)  \right]  \frac{1}{m-\hat{q}_{1}}%
\Gamma_{A^{\prime}A}^{(0)}\,,
\end{equation}
On the other hand, the same expression can be found in the one--loop
approximation using the $s_{i}$--channel unitarity conditions
\cite{Bogdan:2006af,FS}. Comparing these two approaches one obtains%
\begin{align}
&  \left(  \mathcal{R}^{(0)}-\mathcal{L}^{(0)}\right)  \delta_{1}-\delta
_{2}\left(  \mathcal{R}^{(0)}-\mathcal{L}^{(0)}\right) \nonumber\\
&  =\gamma_{\text{Born}}^{G}(q_{1},q_{2})\delta_{1}+\delta_{2}\gamma
_{\text{Born}}^{G}(q_{1},q_{2})\nonumber\\
&  +g^{2}(\hat{q}_{2\bot}-m)\left\{  N\int\frac{\mathtt{d}^{D-2}r_{\bot}%
}{(2\pi)^{D-1}}\,\,gt^{G}e_{G}^{\ast\mu}C^{\mu}(q_{1}+r,q_{2}+r)\frac
{\,\hat{r}_{\bot}-m\,}{(r+q_{2})_{\bot}^{2}(r_{\bot}^{2}-m^{2})(r+q_{1}%
)_{\bot}^{2}}\right. \\
&  +\left.  \frac{1}{N}\int\frac{\mathtt{d}^{D-2}r_{\bot}}{(2\pi)^{D-1}%
}\,\,\frac{\left[  (\hat{q}_{2}+\hat{r})_{\bot}+m\right]  \,\gamma
_{\text{Born}}^{G}(q_{1}+r,q_{2}+r)\left[  \,(\hat{q}_{1}+\hat{r})_{\bot
}\,+m\right]  }{\left[  (r+q_{2})_{\bot}^{2}-m^{2}\right]  \,r_{\bot}%
^{2}\,\left[  (r+q_{1})_{\bot}^{2}-m^{2}\right]  }\right\}  \,(\hat{q}_{1\bot
}-m),\nonumber
\end{align}
which gives for the massless case%
\begin{align}
&  \mathcal{R}^{(0)}-\mathcal{L}^{(0)}=\gamma_{\text{Born}}^{G}(q_{1}%
,q_{2})\frac{\delta_{1}+\delta_{2}}{\delta_{1}-\delta_{2}}\nonumber\\
&  +\frac{g^{2}N}{\delta_{1}-\delta_{2}}\int\frac{\mathtt{d}^{D-2}r_{\bot}%
}{(2\pi)^{D-1}}\,\,gt^{G}e_{G}^{\ast\mu}C^{\mu}(q_{1}+r,q_{2}+r)\frac{\hat
{q}_{2\bot}\,\hat{r}_{\bot}\,\hat{q}_{1\bot}}{r_{\bot}^{2}(r+q_{1})_{\bot}%
^{2}(r+q_{2})_{\bot}^{2}}\label{R-L integral}\\
&  +\frac{g^{2}}{\delta_{1}-\delta_{2}}\frac{1}{N}\int\frac{\mathtt{d}%
^{D-2}r_{\bot}}{(2\pi)^{D-1}}\,\,\frac{\hat{q}_{2\bot}(\hat{q}_{2}+\hat
{r})_{\bot}\,\gamma_{\text{Born}}^{G}(q_{1}+r,q_{2}+r)\,(\hat{q}_{1}+\hat
{r})_{\bot}\,\hat{q}_{1\bot}}{r_{\bot}^{2}(r+q_{1})_{\bot}^{2}(r+q_{2})_{\bot
}^{2}}\,,\nonumber
\end{align}
where
\begin{align}
C^{\mu}(q_{1},q_{2})  &  =-(q_{1}+q_{2})_{\bot}^{\mu}+p_{A}^{\mu}\left(
\frac{s_{2}}{s}+\frac{2q_{1\bot}^{2}}{s_{1}}\right)  -p_{B}^{\mu}\left(
\frac{s_{1}}{s}+\frac{2q_{2\bot}^{2}}{s_{2}}\right) \nonumber\\
&  =-(q_{1}+q_{2})^{\mu}+2\frac{p_{A}^{\mu}}{s_{1}}\left(  \mathbf{k}%
^{2}+t_{1}\right)  -2\frac{p_{B}^{\mu}}{s_{2}}\left(  \mathbf{k}^{2}%
+t_{2}\right)  \,. \label{C born}%
\end{align}
The integral over the transverse components of the virtual gluon momentum
$r^{\mu}$ is calculated in Appendix \ref{appendix A}. Then, substituting the
orthogonal momenta\textbf{ }%
\begin{align}
q_{1\bot}^{\mu}  &  =q_{1}^{\mu}-\frac{s_{2}-t_{1}}{s}p_{A}^{\mu}%
\,-\frac{t_{1}}{s}p_{B}^{\mu}\simeq q_{1}^{\mu}-\frac{s_{2}}{s}p_{A}^{\mu
},\nonumber\\
q_{2\bot}^{\mu}  &  =q_{2}^{\mu}+\frac{s_{1}-t_{2}}{s}p_{B}^{\mu}+\frac{t_{2}%
}{s}p_{A}^{\mu}\,\simeq q_{2}^{\mu}+\frac{s_{1}}{s}p_{B}^{\mu}%
\end{align}
we obtain%
\begin{align}
&  \mathcal{R}^{(0)}-\mathcal{L}^{(0)}=\frac{\delta_{1}+\delta_{2}}{\delta
_{1}-\delta_{2}}\gamma_{\text{Born}}^{G}+\frac{1}{N}\frac{\gamma_{\text{Born}%
}^{G}}{\delta_{1}-\delta_{2}}\frac{2g^{2}\Gamma(1+\epsilon)}{(4\pi)^{D/2}%
}\left(  \frac{1}{\epsilon}+\ln\left(  \frac{\mathbf{k}^{2}}{\mathbf{q}%
_{1}^{2}\mathbf{q}_{2}^{2}}\right)  \right) \nonumber\\
&  -4\left(  N-C_{F}\right)  \frac{g^{2}(4\pi)^{-D/2}}{\delta_{1}-\delta_{2}%
}\left\{  \hat{q}_{2}\gamma_{\text{Born}}^{G}\hat{q}_{1}\left[  \ln\left(
\frac{\mathbf{q}_{1}^{2}}{\mathbf{k}^{2}}\right)  \frac{(\mathbf{kq}_{2}%
)}{\mathbf{k}^{2}t_{2}}-\ln\left(  \frac{\mathbf{q}_{2}^{2}}{\mathbf{k}^{2}%
}\right)  \frac{(\mathbf{kq}_{1})}{\mathbf{k}^{2}t_{1}}\right]  \right.
\nonumber\\
&  -\,gt^{G}e_{G}^{\ast\mu}C^{\mu}(q_{1},q_{2})\left(  \frac{\hat{q}_{1}%
}{t_{1}}-\frac{\hat{q}_{2}}{t_{2}}\right)  \frac{1}{2\mathbf{k}^{2}}\left[
\ln\left(  \frac{\mathbf{q}_{1}^{2}}{\mathbf{k}^{2}}\right)  t_{1}-\ln\left(
\frac{\mathbf{q}_{2}^{2}}{\mathbf{k}^{2}}\right)  t_{2}\right]
\label{R-L answer 1}\\
&  +\left.  \,gt^{G}\left(  \frac{(p_{A}e_{G}^{\ast})}{s_{1}}-\frac
{(p_{B}e_{G}^{\ast})}{s_{2}}\right)  \frac{\hat{k}}{\mathbf{k}^{2}}\left[
\ln\left(  \frac{\mathbf{q}_{1}^{2}}{\mathbf{k}^{2}}\right)  t_{1}-\ln\left(
\frac{\mathbf{q}_{2}^{2}}{\mathbf{k}^{2}}\right)  t_{2}\right]  \right\}
\,.\nonumber
\end{align}
\qquad\qquad\qquad\qquad\qquad

As for the sum $\mathcal{R}^{(1)}+\mathcal{L}^{(1)},$ it is clear that,
contrary to the difference $\mathcal{R}^{(0)}-\mathcal{L}^{(0)}$, it can not
be defined by means of $s_{i}$--channel unitarity in the multi--Regge region.
To calculate it we use a more suitable here $t$--channel dispersive approach
based on analyticity and $t$--channel unitarity, developed in
\cite{Fadin:1993wh,Fadin:2000yp,Fadin:1992zt}.

\section{t--channel discontinuity\label{t -channel section}}

\begin{figure}[h]
\centering
\includegraphics{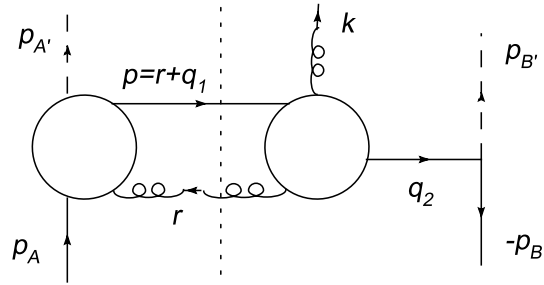}\caption{$t_{1}$--channel discontinuity.}%
\label{fig2}%
\end{figure}

It is possible to calculate the corrections to the RRP vertex by considering
gluon production in various annihilation processes: quark--antiquark,
gluon--gluon, quark--gluon or antiquark--gluon. Moreover, we can consider
photons instead of gluons as the scattered particles, which noticeably reduces
the number of contributing diagrams. Of course, if the Reggeization hypothesis
is valid, we must obtain the same vertex. In the approach based on $t_{i}%
$--channel unitarity it looks very natural.

We choose to consider gluon production in the process of quark--antiquark
annihilation into two photons, therefore in Figs.\ref{fig1}, \ref{fig2} the
particles $A^{\prime}$, $B^{\prime}$ are photons and $G$ is a gluon. Let us
take into account the amplitude discontinuity in the $t_{1}$ channel. The
discontinuity in the $t_{2}$ channel could be considered analogously. The
contribution of $t_{1}$ channel is represented schematically in Fig.\ref{fig2}%
. It is given by an ordinary Cutkosky rule from cut of the corresponding
diagrams.\ After this cut we come to consider the amplitudes being the left
part ($\mathcal{A}$) and the right part ($\mathcal{B}$) of Fig.\ref{fig2}.
Since the external gluon is physical and the cut lines in Fig. \ref{fig2}\ are
on the mass shell, both $\mathcal{A}$ and $\mathcal{B}$ are invariant under
gauge transformations of the intermediate gluon polarization. Therefore, one
can use an arbitrary gauge. We choose Feynman gauge with the polarization
tensor $-g^{\mu\nu}$.

After the convolution we substitute $\mathtt{i}(p^{2}-m^{2}+\mathtt{i}o)^{-1}$
for the on--mass--shell $\delta$--functions $2\pi\,\delta_{+}(p^{2}-m^{2})$
and perform the loop integration thus restoring the amplitude $A_{2\rightarrow
3}^{t_{1}}$. This amplitude has the same $t_{1}$--channel singularities as the
complete one. That is enough to restore the correct Regge asymptotic.\textbf{
}Indeed, obtained amplitude has the same $t_{1}$ singularity as the real one
and an arbitrary polynomial in $t_{1}$ could be added to the final result in
the framework of this procedure but, such term would have a wrong asymptotic
behavior incompatible with the renormalizability of the theory \cite{BFKL},
and, contrary to the case with massive quarks, where the expression which is
equal to the Born amplitude with some constant coefficient can be added, for
the masless limit considered in this paper in detail the correct analytical
properties together with the unitarity requirement in $t_{i}$--channels
determine the amplitude in an unambiguous way.

In order to construct the amplitude with true Regge behavior in MRK, we have
to ensure that it has correct quark quantum numbers, i.e. color triplet and
positive signature in $t_{i}$ channels. While the former is done by applying
the projection operator (which is not actually necessary when $A^{\prime}$ and
$B^{\prime}$ are photons)%
\begin{equation}
\langle c_{1}|\hat{\mathcal{P}}_{3}|c_{2}\rangle=\frac{t^{c_{1}}t^{c_{2}}%
}{C_{F}}, \label{P_quark}%
\end{equation}
the latter can be achieved via the "signaturization" procedure, which is
naturally formulated for truncated amplitudes --- the amplitudes with omitted
wave functions (polarization vectors and Dirac spinors). The procedure is most
simple for the massless case --- to obtain positive signature states in
$t_{1}$ and $t_{2}$ channels we have to symmetrize the truncated amplitude
$A_{2\rightarrow3}^{t_{1}}$\ with respect to the substitutions%
\begin{equation}
p_{A}\rightarrow-p_{A^{\prime}}\,,\quad s_{1}\rightarrow-s_{1}%
\,,\,\,s\rightarrow-s\,,
\end{equation}
and
\begin{equation}
p_{B}\rightarrow-p_{B^{\prime}}\,,\quad s_{2}\rightarrow-s_{2}%
\,,\,\,s\rightarrow-s\,.\,
\end{equation}

So, we have to consider
\begin{equation}
A_{2\rightarrow3}^{t_{1}}=\hat{\mathcal{S}}_{+}\,\,{\int}\frac{\mathtt{d}%
^{D}r}{\mathtt{i(}2\pi\mathtt{)}^{D}}\frac{\sum\mathcal{B}\,\mathcal{A}%
}{\left(  p^{2}-m^{2}+\mathtt{i}o\right)  \left(  r^{2}+\mathtt{i}o\right)
}~,\quad\ p=q_{1}+r\,. \label{BA int}%
\end{equation}
where the sum is taken over all polarizations of the intermediate particles,
the convolution is performed on--mass--shell ($r^{2}=0$, $p^{2}=m^{2}$), and
$\hat{\mathcal{S}_{+}}$ stands for the "signaturization" operator returning
the amplitude with positive signature.

On the one hand, we must use the exact expression for the amplitude
$\mathcal{A}$, as we integrate over the momenta $r$ and $p$ of the
intermediate particles. On the other hand, since $s_{2}\mathcal{\simeq}%
(p_{B}+k)^{2}$ is fixed and large, we take the amplitude $\mathcal{B}$ in the
quasi--multi--Regge kinematics, which means that gluon $G$ with momentum $k$
is produced in the fragmentation region of the intermediate quark and gluon:%
\begin{equation}
\mathcal{B}=\bar{\Gamma}_{B^{\prime}B}^{(0)}\frac{1}{m-\hat{q}_{2}}%
\Gamma_{\{r,k\}\,p}\,.
\end{equation}
Here $\bar{\Gamma}_{B^{\prime}B}^{(0)}=-e\bar{\upsilon}_{B}\hat{e}_{B^{\prime
}}^{\ast}$ and $e$ is a quark--photon coupling constant. We use the
light--cone gauge for the external photon $(e_{B^{\prime}}p_{B^{\prime}%
})=(e_{B^{\prime}}p_{A})=0.$ One can show that the gauge invariant vertex
$\Gamma_{\{r,k\}\,p}$, initially calculated from an inelastic quark--gluon
process \cite{Lipatov:2000se}, does not depend on the nature of the particles
$B$ and $B^{\prime}$, so $B^{\prime}$ can be a photon. This vertex has the
form%
\begin{align}
\Gamma_{\{k_{1},k_{2}\}\,p}  &  =-g^{2}\left\{  t^{2}t^{1}\,\gamma^{\nu}%
(k_{2})\frac{1}{\hat{p}-\hat{k}_{1}-m}\gamma^{\mu}+t^{1}t^{2}\,\gamma^{\mu
}(k_{1})\frac{1}{\hat{p}-\hat{k}_{2}-m}\gamma^{\nu}\right. \nonumber\\
&  \left.  +\frac{\left[  t^{2},t^{1}\right]  }{\left(  k_{1}+k_{2}\right)
^{2}}\gamma^{\rho}(k_{1}+k_{2})\gamma^{\mu\nu\rho}(k_{1},k_{2})+(\hat{q}%
_{2}-m)\frac{p_{B}^{\mu}p_{B}^{\nu}}{\left(  k_{1}+k_{2},p_{B}\right)
}\left(  \frac{t^{1}t^{2}}{\left(  k_{1}p_{B}\right)  }+\frac{t^{2}t^{1}%
}{\left(  k_{2}p_{B}\right)  }\right)  \right\}  u_{p}e_{1}^{\ast\mu}%
e_{2}^{\ast\nu}\,,
\end{align}
where
\begin{align}
\gamma^{\mu}(r)  &  =\gamma^{\mu}+(\hat{q}_{2}-m)\frac{p_{B}^{\mu}}{\left(
rp_{B}\right)  }\,,\quad(k_{i}e_{i}^{\ast})=0,\nonumber\\
\gamma^{\mu\nu\rho}(k_{1},k_{2})  &  =(k_{1}-k_{2})^{\rho}g^{\mu\nu}%
-2k_{1}^{\nu}g^{\rho\mu}+2k_{2}^{\mu}g^{\rho\nu}.
\end{align}
So, projecting the color state of $\mathcal{B}$ on quark quantum numbers
(\ref{P_quark}), we get for massless quarks%
\begin{align}
\Gamma_{\{r,k\}\,p}  &  =-g^{2}\frac{t^{G}t^{r}}{C_{F}}\left\{  C_{F}\left(
\hat{e}_{G}^{\ast}+\hat{q}_{2}\frac{2\left(  e_{G}^{\ast}p_{B}\right)  }%
{s_{2}}\right)  \frac{1}{\hat{q}_{1}}\hat{e}_{r}^{\ast}\,-\frac{1}{2N}\left(
\hat{e}_{r}^{\ast}-\hat{q}_{2}\frac{2\left(  e_{r}^{\ast}p_{B}\right)  }%
{s_{B}}\right)  \frac{1}{\hat{p}-\hat{k}}\hat{e}_{G}^{\ast}\right. \nonumber\\
&  +\frac{N}{2}\frac{1}{l}\left(  \left[  \,\hat{r}-\hat{k}+\hat{q}_{2}\left(
1-2\frac{s_{2}}{u_{B}}\right)  \right]  (e_{r}^{\ast}e_{G}^{\ast})+2\left[
\hat{e}_{G}^{\ast}+\hat{q}_{2}\frac{2(e_{G}^{\ast}p_{B})}{u_{B}}\right]
(ke_{r}^{\ast})\right. \label{G-B}\\
&  -\left.  2\left[  \hat{e}_{r}^{\ast}+\hat{q}_{2}\frac{2(e_{r}^{\ast}p_{B}%
)}{u_{B}}\right]  (re_{G}^{\ast})\right)  +\left.  \hat{q}_{2}\frac
{2(e_{r}^{\ast}p_{B})(e_{G}^{\ast}p_{B})}{s_{2}}\left(  \frac{N}{u_{B}}%
+\frac{1}{N}\frac{1}{s_{B}}\right)  \right\}  u_{p}\,.\nonumber
\end{align}
Here we use some of the next denotations
\begin{align}
(p_{A}+r)^{2}  &  =s_{A}\,,\quad\left(  p_{A}-p\right)  ^{2}=u_{A}%
\,,\nonumber\\
\left(  p_{B}-r\right)  ^{2}  &  =s_{B}\,,\quad(p_{B}+p)^{2}=u_{B}\,,\\
\left(  r+k\right)  ^{2}  &  =l\,,\quad(p-k)^{2}=n\,,\nonumber
\end{align}
and the properties of our kinematics, obtained for the mass shell momenta $r$,
$p$, and $k$:
\begin{align}
s_{A}+u_{A}  &  =-t_{1}\,,\quad s_{B}+u_{B}\simeq s_{2}\,,\quad l+n=t_{2}%
-t_{1}\,,\nonumber\\
2(rp_{A})  &  =s_{A}\,\,,\quad2(pp_{A})=-u_{A}\,\,,\quad2(rp_{B})=-s_{B}\,,\\
2(pp_{B})  &  =u_{B}\,,\quad2(q_{1}p_{B})\simeq s_{2}\,,\quad2(q_{2}%
p_{A})\simeq-s_{1}.\nonumber
\end{align}

As for the amplitude $\mathcal{A}$, it is very profitable here to decompose it
into an "asymptotic" ("as") part, giving the asymptotic of the amplitude
$\mathcal{A}$ in the Regge limit $|s_{A}|\approx|u_{A}|\gg|t_{1}|$, and a
"non--asymptotic" ("na") part:%
\begin{align}
\mathcal{A}  &  =\mathcal{A}^{(as)}+\mathcal{A}^{(na)}\,,\nonumber\\
\mathcal{A}^{(as)}  &  =\bar{\Gamma}_{p\,r}^{(as)}\frac{1}{m-\hat{q}_{1}%
}\Gamma_{A^{\prime}A}^{(0)}\,,\quad\mathcal{A}^{(na)}=-egt^{r}\bar{u}_{p}%
\frac{\hat{e}_{A^{\prime}}^{\ast}\hat{r}\,\hat{e}_{r}}{s_{A}-m^{2}}u_{A}\,,
\label{as non-as}%
\end{align}
where
\begin{equation}
\Gamma_{A^{\prime}A}^{(0)}=-e\,\hat{e}_{A^{\prime}}^{\ast}u_{A}\,,\quad
\bar{\Gamma}_{p\,r}^{(as)}=-gt^{r}\bar{u}_{p}\left(  \hat{e}_{r}+(\hat{q}%
_{1}-m)\frac{2\left(  p_{A}e_{r}\right)  }{s_{A}-m^{2}}\right)  \,.
\label{G as}%
\end{equation}
These parts are invariant under gauge transformations of the intermediate
gluon polarization as well as the complete amplitude. The "asymptotic"
contribution to the full amplitude for massless quarks takes the form:%
\begin{equation}
A_{2\rightarrow3}^{t_{1}(as)}=\hat{\mathcal{S}}_{+}\,\bar{\Gamma}_{B^{\prime
}B}^{(0)}\frac{1}{\hat{q}_{2}}{\int}\frac{\mathtt{d}^{D}r}{\mathtt{i}%
(2\pi)^{D}}\frac{\sum\Gamma_{\{r,k\}p}\,\bar{\Gamma}_{p\,r}^{(as)}}{r^{2}%
p^{2}}\frac{1}{\hat{q}_{1}}\Gamma_{A^{\prime}A}^{(0)}\,. \label{A2-3(as)}%
\end{equation}
The convolution of the vertices yields:
\begin{align}
\sum\Gamma_{\{r,k\}p}\,\bar{\Gamma}_{p\,r}^{(as)}  &  =g^{2}C_{F}\left(
\gamma_{\text{Born}}^{G}-gt^{G}\,\hat{q}_{1}\frac{2(p_{A}e_{G}^{\ast}%
\,)}{s_{1}}\right)  \frac{1}{\hat{q}_{1}}\left(  \gamma^{\mu}\hat{p}%
\gamma^{\mu}+2\frac{\hat{p}_{A\,}\hat{p}\,\hat{q}_{1}}{s_{A}}\right)
\nonumber\\
&  +g^{3}t^{G}\left[  \frac{1}{N}\left(  V_{1}+V_{2}\right)  +N\,\left(
V_{3}+V_{4}\right)  \,\right]  \,, \label{Gamma convolve}%
\end{align}
where%
\begin{align}
V_{1}  &  =\frac{\gamma^{\mu}(\hat{p}-\hat{k})\hat{e}_{G}^{\ast}\hat{p}%
\gamma^{\mu}}{2n}-\frac{\hat{q}_{2}(\hat{p}-\hat{k})\,\hat{e}_{G}^{\ast}%
\hat{p}\,\hat{q}_{1}s}{ns_{A}s_{B}}\nonumber\\
&  +\left(  \frac{\hat{p}_{A\,}(\hat{p}-\hat{k})\,\hat{e}_{G}^{\ast}\hat
{p}\,\hat{q}_{1}}{ns_{A}}-\frac{\hat{q}_{2}(\hat{p}-\hat{k})\,\hat{e}%
_{G}^{\ast}\hat{p}\,\hat{p}_{B}}{ns_{B}}\right)  \,, \label{V1}%
\end{align}%
\begin{equation}
V_{2}=-\frac{2(p_{B}e_{G}^{\ast})}{s_{2}s_{B}}\left(  \hat{q}_{2}\hat{p}%
\hat{p}_{B}+\frac{\hat{q}_{2}\hat{p}\,\hat{q}_{1}s}{s_{A}}\right)  \,,\quad
V_{3}=V_{2}(s_{B}\rightarrow u_{B})\,, \label{V2-V3}%
\end{equation}%
\begin{align}
V_{4}  &  =\frac{\hat{k}\hat{p}\hat{e}_{G}^{\ast}-\hat{e}_{G}^{\ast}\hat
{p}\hat{k}}{l}+\left(  \frac{\hat{k}\hat{p}\,\hat{q}_{1}2(p_{A}e_{G}^{\ast}%
)}{ls_{A}}-\frac{\hat{q}_{2}\hat{p}\,\hat{k}2(p_{B}e_{G}^{\ast})}{lu_{B}%
}\right) \nonumber\\
&  +2\frac{\hat{q}_{2}\hat{p}\,\hat{q}_{1}}{ls_{A}u_{B}}\left(  (p_{A}%
e_{G}^{\ast})s_{2}-(p_{B}e_{G}^{\ast})s_{1}\right)  +\left(  \frac{\hat{q}%
_{2}\hat{p}\,\hat{e}_{G}^{\ast}s_{2}}{lu_{B}}-\frac{\hat{e}_{G}^{\ast}\hat
{p}\,\hat{q}_{1}s_{1}}{ls_{A}}\right) \label{V4}\\
&  +2\left(  \frac{\hat{p}_{A}\hat{p}\,\hat{q}_{1}(r\,e_{G}^{\ast})}{ls_{A}%
}+\frac{\hat{q}_{2}\hat{p}\,\hat{p}_{B}(r\,e_{G}^{\ast})}{lu_{B}}\right)
+\frac{\gamma^{\mu}\hat{p}\gamma^{\mu}(r\hat{e}_{G}^{\ast})}{l}+\frac{\hat
{q}_{2}\hat{p}\,\hat{q}_{1}2(r\,e_{G}^{\ast})s}{ls_{A}u_{B}}\,.\nonumber
\end{align}
The integrals appearing in eq.(\ref{A2-3(as)}) from the convolution
(\ref{Gamma convolve}) are calculated in Appendix \ref{appendix B} of this
paper and Appendix C of \cite{Fadin:1993wh}. Due to its length we do not
present the integrated expression for (\ref{A2-3(as)}) \ here and will use it
for calculating the answer for $\mathcal{R}+\mathcal{L}$ in Section
\ref{answer section}. It is important to stress that we integrate in the fixed
order of limit taking\textbf{,} as it is done systematically in Regge
approach: first $s_{i}\rightarrow\infty$, and only after it $D\rightarrow4$.
Of course, these two limits must commute in final infrared stable results for
observables, but at intermediate steps one must adhere to the initially set
order everywhere. The consequence of our choice is the prohibition to expand
such terms as $s^{-\epsilon}$\ to series with respect to $\epsilon
\rightarrow0$.

The integrated expression for (\ref{A2-3(as)}) has a correct singularity
(discontinuity) in the $t_{1}$ channel. It also has a $t_{2}$ channel
singularity, but the latter is correct only for the terms with both $t_{1}$
and $t_{2}$ channel singularity. The amplitude with correct analytic
properties in both channels can be easily yielded by symmetry consideration.
Using the uncertainty of the dispersion formula (\ref{A2-3(as)}) with respect
to adding to the convolution (\ref{Gamma convolve}) some terms proportional to
$r^{2}$ or $p^{2}$, we can restore the expression which is symmetric (with
respect to change left and right parts of the diagram (see. fig. \ref{fig1})),
and vanishes after $e_{G}\rightarrow k$ substitution. We find such additional
terms after analyzing \ the integrated form of (\ref{A2-3(as)}). They depend
on how we treat hidden $r^{2}$ and $p^{2}$ in $V$'s. In all cases we commute
terms similar to $\hat{r}\hat{e}_{G}^{\ast}\hat{r}$ into 2$\hat{r}(e_{G}%
^{\ast}r)-r^{2}\hat{e}_{G}^{\ast}$ and set $r^{2}$ to 0. So our terms are
\begin{align}
&  p^{2}\hat{\mathcal{S}}_{1\leftrightarrow2}\frac{1}{p^{2}}\left\{
g^{2}C_{F}\left(  \gamma_{\text{Born}}^{G}-gt^{G}\,\hat{q}_{1}\frac
{2(p_{A}e_{G}^{\ast}\,)}{s_{1}}\right)  \frac{1}{\hat{q}_{1}}\left(
\gamma^{\mu}\hat{p}\gamma^{\mu}+2\frac{\hat{p}_{A\,}\hat{p}\,\hat{q}_{1}%
}{s_{A}}\right)  +g^{3}t^{G}\left(  \frac{1}{N}V_{2}+NV_{3}\right)  \right\}
\nonumber\\
&  +\frac{r^{2}}{n}\frac{g^{3}t^{G}}{N}\left(  \frac{s\hat{q}_{2}\hat{e}%
_{G}^{\ast}\hat{q}_{1}}{s_{A}s_{B}}+\frac{2\hat{q}_{2}(p_{B}e_{G}^{\ast}%
)}{s_{B}}-\frac{2\hat{q}_{1}(p_{A}e_{G}^{\ast})}{s_{A}}\right)  +\frac{p^{2}%
}{n}N\frac{g^{3}t^{G}}{s_{B}}\left\{  2\hat{q}_{2}\hat{p}_{B}\hat{e}_{G}%
^{\ast}+\hat{e}_{G}^{\ast}t_{2}+\hat{q}_{2}\hat{p}_{B}\hat{q}_{2}\frac
{2(p_{B}e_{G}^{\ast})}{s_{2}}\right\}  \,, \label{symmetr add}%
\end{align}
where $\hat{\mathcal{S}}_{1\leftrightarrow2}$ is the operator which flips
gamma matrix order and makes exchange $q_{1}\leftrightarrow-q_{2}$ and
$p_{A}\leftrightarrow p_{B}$ thus yielding the corresponding terms from the
contribution of the $t_{2}$ channel discontinuity. Expression
(\ref{symmetr add}) is applicable for arbitrary $D$. After adding
eq.(\ref{symmetr add}) to (\ref{Gamma convolve}) we do not need to consider
the contribution of the amplitude discontinuity in the $t_{2}$ channel.

It is clear that regardless of energy normalization, $\mathcal{R}%
^{(1)}+\mathcal{L}^{(1)}$ contributes only to the real part of the l.h.s. of
eq.(\ref{A one-loop}) and $\mathcal{R}^{(0)}-\mathcal{L}^{(0)}$ is
unambiguously determined by its imaginary part. Since both $\gamma
_{\text{Born}}^{G}$ \ and $A_{2\rightarrow3}$ vanish after the substitution of
the gluon momentum $k$\ for the polarization vector $e_{G}$, we can conclude
that both $\mathcal{R}^{(1)}+\mathcal{L}^{(1)}$ and \ $\mathcal{R}%
^{(0)}-\mathcal{L}^{(0)}$ vanish as well.

Now, let us turn to the "non--asymptotic" contribution $A_{2\rightarrow
3}^{t_{1}(na)}$ given by the product $\mathcal{BA}_{na}$. Since the
"non--asymptotic" part $\mathcal{A}_{na}$ (see (\ref{as non-as})) does not
contain terms of order $s_{A}$ for large values of this invariant, the
essential region of integration over $r$ in this case is%
\begin{align}
&  s_{A}\sim u_{A}\sim r^{2}\sim p^{2}\sim t_{i}\,,\nonumber\\
l  &  \sim n\sim s_{1}\,,\quad s_{B}\sim u_{B}\gg|t_{i}|\,. \label{na region}%
\end{align}
It implies that, in order to calculate the contribution of $A_{2\rightarrow
3}^{t_{1}(na)}$, one may take the amplitude $\mathcal{B}$ in the multi--Regge
asymptotic region instead of complicated expression (\ref{G-B}). In that
region one may use the simple multi--Regge form (\ref{A-LLA}) with the
corresponding substitution of momenta and color indices. Therefore, we obtain
\begin{align}
\mathcal{B}(\text{"na"--region})  &  =\bar{\Gamma}_{B^{\prime}B}^{(0)}\frac
{1}{m-\hat{q}_{2}}\gamma_{\text{Born}}^{G}\frac{1}{m-\hat{q}_{1}}\Gamma
_{rp}^{k(as)}\,,\nonumber\\
\Gamma_{rp}^{k(as)}  &  =-gt^{r}\left(  \hat{e}_{r}^{\ast}+(\hat{q}%
_{1}-m)\frac{\left(  ke_{r}^{\ast}\right)  }{(k\,r)}\right)  u_{p}\,.
\label{B na}%
\end{align}
Thus, the problem of calculating the $A_{2\rightarrow3}^{t_{1}(na)}$
contribution reduces to a simpler problem for the elastic case. Moreover, it
is not necessary to calculate this contribution at all, since it is totally
absorbed in the correction to the PPR vertices $\Gamma_{P^{\prime}P}^{(1)}$ .
We stress that one--loop corrections to the gluon production amplitude
$A_{2\rightarrow3}($one-loop$)$ include corrections to these vertices, as one
can see from eq.(\ref{A one-loop}). We will clarify this point in the next Section.

\section{Asymptotic correction to PPR vertex}

\begin{figure}[h]
\centering
\includegraphics{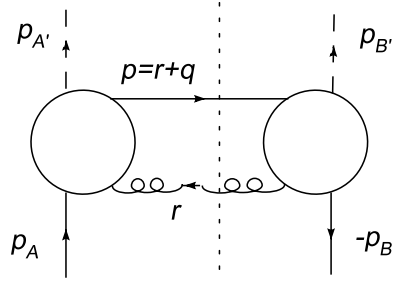}\caption{Calculation of $\Gamma_{P^{\prime}P}%
^{(1)}.$}%
\label{fig3}%
\end{figure}The calculation of $\mathcal{R}^{(1)}\mathcal{+L}^{(1)}$ includes
the one loop correction to the PPR vertex $\Gamma_{P^{\prime}P}^{(1)}$ (see
eq.(\ref{A one-loop})). Assuming the validity of the Reggeization hypothesis,
one may find $\Gamma_{P^{\prime}P}^{(1)}$ from cosideration of the amplitude
of quark--antiquark annihilation into photons in Regge limit. We compare the
projection of this amplitude on the positive signature and its Regge form:%

\begin{equation}
A_{2\rightarrow2}=\frac{1}{2}\bar{\Gamma}_{B^{\prime}B}(q,s_{0})\frac
{1}{m-\hat{q}}\left[  \left(  \frac{s}{s_{0}}\right)  ^{\delta(q)}+\left(
\frac{-s}{s_{0}}\right)  ^{\delta(q)}\right]  \Gamma_{A^{\prime}A}(q,s_{0})\,,
\label{A2-2}%
\end{equation}
where $s_{0}$ is an energy normalization point. Eq. (\ref{A-NLA}) is written
for PPR vertex $\Gamma_{P^{\prime}P}$ normalized at $s_{0}=-t$, so we
calculate the correction to this vertex at the same point. The one--loop
approximation gives us%
\begin{equation}
A_{2\rightarrow2}(\text{one--loop})=\bar{\Gamma}_{B^{\prime}B}^{(1)}\frac
{1}{m-\hat{q}}\Gamma_{A^{\prime}A}^{(0)}+\bar{\Gamma}_{B^{\prime}B}^{(0)}%
\frac{1}{m-\hat{q}}\Gamma_{A^{\prime}A}^{(1)}+\frac{1}{2}\bar{\Gamma
}_{B^{\prime}B}^{(0)}\frac{\delta(q)}{m-\hat{q}}\Gamma_{A^{\prime}A}^{(0)}%
\ln\left(  \frac{s(-s)}{(-t)^{2}}\right)  \,. \label{A2-2 one-loop}%
\end{equation}
Let us note that formula (\ref{A2-2 one-loop}), i.e. the Reggeization
hypothesis\ at $\alpha_{s}^{2}$ order, is proved \cite{BD-DFG}.

Similarly to the previous Section we use $t$ channel unitarity technique and,
performing the Cutkosky cut, deal with the left ($\mathcal{A}$) and the right
($\mathcal{B}$) parts of Fig.\ref{fig3}. After $q_{1}\rightarrow q$
substitution the amplitude $\mathcal{A}$ coincides with the ones presented in
(\ref{as non-as}) and the amplitude $\mathcal{B}$ can be easily obtained by
its Hermitian conjugation and $p_{A}\leftrightarrow-p_{B}$ exchange. We mark
its parts with letter $B$ (e.g. $\Gamma_{p\,r}^{B(as)}$).

The contribution of the \ product $\mathcal{B}\,^{(as)}\mathcal{A}^{(na)}$
represents a part of the corrections to the amplitude connected with the piece
of $\ \bar{\Gamma}_{A^{\prime}A}^{(1)}$. Because of the factorization property
of $\mathcal{B}^{(as)}$ we have for the massless case%
\begin{equation}
\Gamma_{A^{\prime}A}^{(1)(as-na)}=\hat{\mathcal{S}}_{+}%
%TCIMACRO{\dint }%
%BeginExpansion
{\displaystyle\int}
%EndExpansion
\frac{\mathtt{d}^{D}r}{\mathtt{i(}2\pi\mathtt{)}^{D}}\frac{\sum\Gamma
_{rp}^{B(as)}\mathcal{A}^{(na)}}{r^{2}p^{2}}\,, \label{G as-na}%
\end{equation}
Here, as in the previous Section we use the light--cone gauges for external
photons
\begin{equation}
(e_{A^{\prime}}p_{A^{\prime}})=(e_{B^{\prime}}p_{B^{\prime}})=(e_{A^{\prime}%
}p_{B})=(e_{B^{\prime}}p_{A})=0\,, \label{photon gauge}%
\end{equation}
which means, in other words, that the final result in general gauges will be
given by the replacements%
\[
e_{A^{\prime}}\rightarrow e_{A^{\prime}}-p_{A^{\prime}}\frac{(e_{A^{\prime}%
}p_{B})}{(p_{A^{\prime}}p_{B})}\,,\quad e_{B^{\prime}}\rightarrow
e_{B^{\prime}}-p_{B^{\prime}}\frac{(e_{B^{\prime}}p_{A})}{(p_{A}p_{B^{\prime}%
})}\,.
\]
The integrand of (\ref{G as-na}) is
\begin{equation}
\sum\Gamma_{rp}^{B(as)}\mathcal{A}^{(na)}=-eg^{2}C_{F}\left(  \frac
{\gamma^{\mu}\,\hat{p}\,\hat{e}_{A^{\prime}}\hat{r}\gamma^{\mu}}{s_{A}}%
-2\frac{\hat{q}\,\hat{p}\,\,\hat{e}_{A^{\prime}}\hat{r}\,\hat{p}_{B}}%
{s_{A}s_{B}}\right)  u_{A}\,.
\end{equation}
Necessary integrals can be found in Appendix \ref{appendix C} of this paper
and in Appendix A of \cite{Fadin:1993wh}. We obtain
\begin{equation}
\Gamma_{A^{\prime}A}^{(1)(as-na)}=-e\,\delta(q)\left(  \hat{e}_{A^{\prime}%
}\frac{-3-2\epsilon}{2(1-2\epsilon)}+\frac{\hat{q}(e_{A^{\prime}}q)}{t}%
\frac{2+\epsilon}{1-2\epsilon}\right)  u_{A}\,,
\end{equation}
where $\delta(q)$ is the quark trajectory (\ref{trajectory}) for $m=0$.

While inserting $\Gamma_{A^{\prime}A}^{(1)(as-na)}$ (\ref{G as-na}) into
(\ref{A one-loop}) we substitute $p_{B}\rightarrow k$, which leads to
$\Gamma_{rp}^{B(as)}\rightarrow\Gamma_{rp}^{k(as)}$ (cf.\ eq.(\ref{B na})).
Therefore, due to the analogous factorization property of \ $\mathcal{B}%
($"na"--region$)$ represented in eq.(\ref{B na}), the contribution of \ the
product \ $\mathcal{BA}^{(na)}$ to the gluon production amplitude
$A_{3\rightarrow2}($one--loop$)$ cancels the piece of the second term of
(\ref{A one-loop}) given by $\Gamma_{A^{\prime}A}^{(1)(as-na)}$. So, actually,
we need to calculate neither the contribution of $\mathcal{BA}^{(na)}$ to
$A_{3\rightarrow2}$, nor the contribution of $\mathcal{B}\,^{(as)}%
\mathcal{A}^{(na)}$ to $A_{2\rightarrow2}$. The symmetric situation takes
place for the part of $\bar{\Gamma}_{B^{\prime}B}^{(1)}$.

The totally "non--asymptotic" contribution to the $A_{2\rightarrow2}$ for
massless quarks is given by the product $\mathcal{B}\,^{(na)}\mathcal{A}%
^{(na)}.$ For it, the essential values of $(rp_{A})$ and $(rp_{B})$ in the
integration region are small: $r^{2}\sim p^{2}\sim s_{A}\sim s_{B}\sim t$.
Therefore, in Sudakov decomposition (\ref{Sudakov}) parameters $\alpha$ and
$\beta$ are suppressed $\alpha\sim\beta\sim t/s$ and can be omitted in the
propagators:%
\begin{equation}
r^{2}\rightarrow r_{\bot}^{2}\,,\quad p^{2}\rightarrow(r+q)_{\bot}^{2}\,,\quad
s_{A}\rightarrow r_{\bot}^{2}+s\alpha\,,\quad s_{B}\rightarrow r_{\bot}%
^{2}-s\beta\,\,.
\end{equation}
It leads to the factorization of the corresponding integral (see
(\ref{BA int})) into two blob integrals with respect to $\alpha$ and $\beta,$
which can be taken by residues. The corresponding contribution to the total
amplitude is pure imaginary at high energies and corresponds to the
lowest--order term of the negative signature. Here we are interested in
radiative corrections to the positive signatured amplitude and therefore
contribution of $\mathcal{B}\,^{(na)}\mathcal{A}^{(na)}$ is not important for us.

Therefore, we have to consider only the piece of $\Gamma_{A^{\prime}A}^{(1)}$
(and $\bar{\Gamma}_{B^{\prime}B}^{(1)}$) defined by $\mathcal{B}%
\,^{(as)}\mathcal{A}^{(as)}$ and subtract from the corrections to the gluon
production amplitude $A_{3\rightarrow2}($one--loop$)$ in eq.(\ref{A one-loop})
only the part connected with this piece ($\Gamma_{A^{\prime}A}^{(1)(as-as)}%
$).
\begin{equation}
A_{2\rightarrow2}^{(as-as)}=\hat{\mathcal{S}}_{+}\bar{\Gamma}_{B^{\prime}%
B}^{(0)}\frac{1}{-\hat{q}}\,%
%TCIMACRO{\dint }%
%BeginExpansion
{\displaystyle\int}
%EndExpansion
\frac{\mathtt{d}^{D}r}{\mathtt{i}(2\pi)^{D}}\frac{\sum\,\Gamma_{p\,r}%
^{B(as)}\bar{\Gamma}_{p\,r}^{(as)}}{r^{2}p^{2}}\frac{1}{-\hat{q}}%
\Gamma_{A^{\prime}A}^{(0)}\,.
\end{equation}
The integrand here has the form:%
\begin{equation}
\sum\,\Gamma_{p\,r}^{B(as)}\bar{\Gamma}_{p\,r}^{(as)}=-g^{2}C_{F}\left(
\gamma^{\mu}\hat{p}\gamma^{\mu}+\frac{2\hat{p}_{A}\hat{p}\,\hat{q}}{s_{A}%
}-\frac{2\hat{q}\hat{p}\,\hat{p}_{B}\,}{s_{B}}-2s\frac{\hat{q}\hat{p}\,\hat
{q}}{s_{A}s_{B}}\right)  \,. \label{Gamma integrand}%
\end{equation}
After the integration we obtain%
\begin{equation}
\Gamma_{A^{\prime}A}^{(1)(as-as)}=-e\,\delta(q)\left(  \hat{e}_{A^{\prime}%
}\frac{1}{2}\left[  -\frac{1}{\epsilon}+\frac{3+\epsilon}{2(1-2\epsilon)}%
+\psi(1)+\psi(1+\epsilon)-2\psi(1-\epsilon)\right]  -\frac{\hat{q}%
(e_{A^{\prime}}q)}{t}\frac{2}{1-2\epsilon}\right)  u_{A}\,. \label{G as-as}%
\end{equation}
Finally, we present the full photon--quark--Reggeized quark vertex with photon
polarization satisfying light--cone gauge conditions (\ref{photon gauge}):%
\begin{equation}
\Gamma_{A^{\prime}A}=-e\,\left(  \hat{e}_{A^{\prime}}(1+\delta_{e})+\frac
{\hat{q}(e_{A^{\prime}}q)}{t}\delta_{q}\right)  u_{A}\,,
\end{equation}
where%
\begin{align}
\delta_{e}  &  =\,\delta(q)\frac{1}{2}\left[  -\frac{1}{\epsilon}%
-\frac{3(1+\epsilon)}{2(1-2\epsilon)}+\psi(1)+\psi(1+\epsilon)-2\psi
(1-\epsilon)\right]  +\mathcal{O}(g^{4})\,,\nonumber\\
\delta_{e}  &  =\,\delta(q)\,\frac{\epsilon}{1-2\epsilon}+\mathcal{O}%
(g^{4})\,.
\end{align}
This result coincides with the one, which can be easily obtained from the
calculations of the gluon--quark--Reggeized quark vertex, presented in
V.S.\ Fadin and R. Fiore's \ paper \cite{FF01}. Indeed, taking the coefficient
only before $-1/2N$ in their result and multiplying it by $C_{F}$ corresponds
to\ omitting the triple gluon vertex and corrected color algebra.

\section{Radiative correction to RRP vertex\label{answer section}}

Finally, using the table of integrals presented in the Appendices and
gathering results from previous sections we find the one--loop correction to
the RRP vertex. The answer contains terms proportional to three gamma--matrix
structure $\hat{q}_{2}\hat{e}_{G}^{\ast}\hat{q}_{1}$(e.g. term $\hat{q}%
_{2}\gamma_{\text{Born}}^{G}\hat{q}_{1}$ in (\ref{R-L answer 1})), which in
our kinematics (do not forget about surrounding spinors in the amplitude,
where the RRG vertex has to be inserted) can be replaced by $\hat{q}_{2\bot
}\hat{e}_{G\bot}^{\ast}\hat{q}_{1\bot}$, so we have%
\begin{equation}
\hat{q}_{2}\hat{e}_{G}^{\ast}\hat{q}_{1}=\hat{q}_{1}(e_{G}^{\ast}q_{2})_{\bot
}+\hat{q}_{2}(e_{G}^{\ast}q_{1})_{\bot}-(q_{1}q_{2})_{\bot}\hat{e}_{G}^{\ast
}+\mathrm{i\,}\gamma^{5}\gamma^{d}\varepsilon^{abcd}q_{1\bot}^{a}\,e_{G\bot
}^{\ast b}\,q_{2\bot}^{c}\,,
\end{equation}
with the last term equal to zero. There is also a linear relation between the
parts of the RRP vertex resulting from the fact that $e_{G\bot}$ can be
decomposed into the basis vectors: $q_{1\bot}$, $q_{2\bot}$.%
\begin{align}
\gamma_{\text{Born}}^{G} &  =gt^{G}e_{G}^{\ast\mu}C^{\mu}(q_{1},q_{2})\frac
{1}{\Delta}\left[  \left(  \hat{q}_{1}+\hat{q}_{2}\right)  \mathbf{k}%
^{2}+\left(  \hat{q}_{1}-\hat{q}_{2}\right)  (t_{1}-t_{2})\right]  \nonumber\\
&  +gt^{G}\left(  \frac{(p_{A}e_{G}^{\ast})}{s_{1}}-\frac{(p_{B}e_{G}^{\ast}%
)}{s_{2}}\right)  \frac{2t_{1}t_{2}}{\Delta}\left[  \left(  \frac{\hat{q}_{1}%
}{t_{1}}+\frac{\hat{q}_{2}}{t_{2}}\right)  \mathbf{k}^{2}-\left(  \frac
{\hat{q}_{1}}{t_{1}}-\frac{\hat{q}_{2}}{t_{2}}\right)  \left(  t_{1}%
-t_{2}\right)  \right]  \,,
\end{align}
where $\Delta=4(q_{1}q_{2})_{\bot}^{2}-4t_{1}t_{2}$. We used this relation to
simplify our answer for $\mathcal{R}^{(1)}+\mathcal{L}^{(1)}$.%
\begin{align}
&  \left(  \mathcal{R}^{(0)}-\mathcal{L}^{(0)}\right)  \frac{\delta_{1}%
-\delta_{2}}{c_{\Gamma}g^{2}}=2N\,\gamma_{\text{Born}}^{G}\left(  \frac
{1}{\epsilon}-\ln\left(  \mathbf{k}^{2}\right)  \right)  \nonumber\\
&  +\left(  N-C_{F}\right)  \left\{  \gamma_{\text{Born}}^{G}\left[
\frac{\left(  \mathbf{k}^{2}+t_{1}+t_{2}\right)  ^{2}-2t_{1}\left(
t_{1}+t_{2}\right)  }{\mathbf{k}^{2}t_{1}}\ln\left(  \frac{\mathbf{k}^{2}%
}{-t_{2}}\right)  +\frac{\left(  \mathbf{k}^{2}+t_{1}+t_{2}\right)
^{2}-2t_{2}\left(  t_{1}+t_{2}\right)  }{\mathbf{k}^{2}t_{2}}\ln\left(
\frac{\mathbf{k}^{2}}{-t_{1}}\right)  \right]  \,\right.  \nonumber\\
&  +gt^{G}e_{G}^{\ast\mu}C^{\mu}(q_{1},q_{2})\left[  \frac{\hat{q}_{1}+\hat
{q}_{2}}{t_{1}t_{2}}V^{+}+\frac{\hat{q}_{1}-\hat{q}_{2}}{t_{1}t_{2}}%
\frac{t_{1}+t_{2}}{\mathbf{k}^{2}}V^{-}\right]  \label{R-L fin}\\
&  +\left.  2\,gt^{G}\left(  \frac{(p_{A}e_{G}^{\ast})}{s_{1}}-\frac
{(p_{B}e_{G}^{\ast})}{s_{2}}\right)  \left[  \left(  \frac{\hat{q}_{1}}{t_{1}%
}+\frac{\hat{q}_{2}}{t_{2}}\right)  V^{+}-\left(  \frac{\hat{q}_{1}}{t_{1}%
}-\frac{\hat{q}_{2}}{t_{2}}\right)  \frac{\left(  t_{1}+t_{2}\right)
}{\mathbf{k}^{2}}V^{-}\right]  \right\}  \,,\nonumber
\end{align}
where%
\begin{equation}
V^{+}=\ln\left(  \frac{-t_{1}}{\mathbf{k}^{2}}\right)  t_{1}+\ln\left(
\frac{-t_{2}}{\mathbf{k}^{2}}\right)  t_{2}\,,\quad V^{-}=\ln\left(
\frac{-t_{1}}{\mathbf{k}^{2}}\right)  t_{1}-\ln\left(  \frac{-t_{2}%
}{\mathbf{k}^{2}}\right)  t_{2}\,,
\end{equation}%
\begin{equation}
c_{\Gamma}=\frac{\Gamma(1+\epsilon)\Gamma(1-\epsilon)^{2}}{\Gamma
(1-2\epsilon)(4\pi)^{D/2}}\,,\quad\Li(z)=-\int_{0}^{1}\frac{\,\mathtt{d}x}%
{x}\ln(1-zx)\,\,,
\end{equation}
and the expressions for $\gamma_{\text{Born}}^{G}$ and $C^{\mu}$ are defined
in eqs.(\ref{gamma-Born}), (\ref{C born}).
\begin{align*}
&  \left(  \mathcal{R}^{(1)}+\mathcal{L}^{(1)}\right)  \left(  c_{\Gamma}%
g^{2}\right)  ^{-1}\\
&  =\gamma_{\text{Born}}^{G}\left\{  \frac{N}{2}\left[  -\frac{2}{\epsilon
^{2}}+\frac{2}{\epsilon}\ln\left(  \mathbf{k}^{2}\right)  -\ln^{2}\left(
\mathbf{k}^{2}\right)  -\ln^{2}\left(  \frac{-t_{1}}{-t_{2}}\right)
-\frac{3\mathbf{k}^{2}}{t_{1}-t_{2}}\ln\left(  \frac{-t_{1}}{-t_{2}}\right)
+\frac{2\pi^{2}}{3}\right]  \right.  \\
&  +\left(  N-C_{F}\right)  \left[  \frac{\left(  \mathbf{k}^{2}+t_{1}%
+t_{2}\right)  ^{2}-2t_{1}\left(  t_{1}+t_{2}\right)  }{4\mathbf{k}^{2}t_{1}%
}\ln^{2}\left(  \frac{\mathbf{k}^{2}}{-t_{2}}\right)  +\frac{\left(
\mathbf{k}^{2}+t_{1}+t_{2}\right)  ^{2}-2t_{2}\left(  t_{1}+t_{2}\right)
}{4\mathbf{k}^{2}t_{2}}\ln^{2}\left(  \frac{\mathbf{k}^{2}}{-t_{1}}\right)
\right.  \\
&  +\left.  \left.  \frac{\left(  \mathbf{k}^{2}+t_{2}\right)  ^{2}-t_{1}^{2}%
}{2\mathbf{k}^{2}t_{1}}\Li\left(  1-\frac{t_{1}}{t_{2}}\right)  +\frac{\left(
\mathbf{k}^{2}+t_{1}\right)  ^{2}-t_{2}^{2}}{2\mathbf{k}^{2}t_{2}}\Li\left(
1-\frac{t_{2}}{t_{1}}\right)  -\frac{t_{1}+t_{2}-2\mathbf{k}^{2}}{2\left(
t_{1}-t_{2}\right)  }\ln\left(  \frac{-t_{1}}{-t_{2}}\right)  +1\right]
\right\}
\end{align*}
\vspace{-0.4cm}
\[
\,-gt^{G}e_{G}^{\ast\mu}C^{\mu}(q_{1},q_{2})\frac{1}{2}\left\{  \left(
N-C_{F}\right)  \left[  \frac{\hat{q}_{1}+\hat{q}_{2}}{t_{1}t_{2}}W^{+}%
+\frac{\hat{q}_{1}-\hat{q}_{2}}{t_{1}t_{2}}\frac{\left(  t_{1}+t_{2}\right)
}{\mathbf{k}^{2}}W^{-}\right]  +\frac{3}{N}\frac{\hat{q}_{1}+\hat{q}_{2}%
}{\left(  t_{1}-t_{2}\right)  }\ln\left(  \frac{-t_{1}}{-t_{2}}\right)
\right\}
\]
\vspace{-0.4cm}
\begin{align}
&  -gt^{G}\left(  \frac{(p_{A}e_{G}^{\ast})}{s_{1}}-\frac{(p_{B}e_{G}^{\ast}%
)}{s_{2}}\right)  \left\{  N\left(  \frac{\hat{q}_{1}}{t_{1}}+\frac{\hat
{q}_{2}}{t_{2}}\right)  \frac{3t_{1}t_{2}}{t_{1}-t_{2}}\ln\left(  \frac
{-t_{1}}{-t_{2}}\right)  \right.  \nonumber\\
&  +\left(  N-C_{F}\right)  \left[  \left(  \frac{\hat{q}_{1}}{t_{1}}%
+\frac{\hat{q}_{2}}{t_{2}}\right)  \left(  W^{+}-\mathbf{k}^{2}\right)
\right.  \label{R+L fin}\\
&  -\left.  \left.  \left(  \frac{\hat{q}_{1}}{t_{1}}-\frac{\hat{q}_{2}}%
{t_{2}}\right)  \frac{\left(  t_{1}+t_{2}\right)  }{\mathbf{k}^{2}}\left(
W^{-}+\frac{\left(  \mathbf{k}^{2}\right)  ^{2}}{t_{1}-t_{2}}-\frac
{2t_{1}t_{2}\left(  \mathbf{k}^{2}\right)  ^{2}}{\left(  t_{1}-t_{2}\right)
^{2}\left(  t_{1}+t_{2}\right)  }\ln\left(  \frac{-t_{1}}{-t_{2}}\right)
\right)  \right]  \right\}  \,,\nonumber
\end{align}
where
\begin{align}
W^{+} &  =\frac{1}{2}t_{1}\ln^{2}\left(  \frac{\mathbf{k}^{2}}{-t_{1}}\right)
+\frac{1}{2}\ln^{2}\left(  \frac{\mathbf{k}^{2}}{-t_{2}}\right)  t_{2}%
-\frac{3t_{1}t_{2}}{t_{1}-t_{2}}\ln\left(  \frac{-t_{1}}{-t_{2}}\right)
+t_{1}\Li\left(  1-\frac{t_{2}}{t_{1}}\right)  +t_{2}\Li\left(  1-\frac{t_{1}%
}{t_{2}}\right)  \,,\nonumber\\
W^{-} &  =\frac{1}{2}t_{1}\ln^{2}\left(  \frac{\mathbf{k}^{2}}{-t_{1}}\right)
-\frac{1}{2}\ln^{2}\left(  \frac{\mathbf{k}^{2}}{-t_{2}}\right)  t_{2}%
-\frac{\mathbf{k}^{2}t_{1}t_{2}}{\left(  t_{1}-t_{2}\right)  ^{2}}\ln\left(
\frac{-t_{1}}{-t_{2}}\right)  \\
&  +t_{1}\Li\left(  1-\frac{t_{2}}{t_{1}}\right)  -t_{2}\Li\left(
1-\frac{t_{1}}{t_{2}}\right)  +\frac{2\mathbf{k}^{2}t_{1}t_{2}}{t_{1}%
^{2}-t_{2}^{2}}\,.\nonumber
\end{align}
Note that new spin structures appear in eqs.(\ref{R-L fin},\ref{R+L fin}) in
comparison with the Born effective vertex (\ref{gamma-Born}) and that the
corrections obtained are obviously gauge invariant. It is easy to verify that
poles at $t_{1}=t_{2}$ in eq.(\ref{R+L fin}) cancel. Moreover, one can check
that when $|q_{1\bot}|$ or $|q_{2\bot}|$ approaches zero the coefficients at
each spin structure remain finite.
%Further, the integrals like $I_{A}$ and $I_{B}$ (see Appendix
%\ref{appendix B}) which \ appear in calculations of
%$\mathcal{R}^{(0)}-\mathcal{L}^{(0)}$ and give ($\pm s_{i})^{-\epsilon}$ do not
%cancelled in the final result, which means that considered effective vertex
%correction does depend on the adopted order of the limits taking: first
%$s_{i}\rightarrow\infty$, and then $D\rightarrow4$.

Let us stress that the calculated amplitude $A_{2\rightarrow3}($one-loop$)$
has the very same nontrivial structure in energy variables $s_{i}$ as
predicted by Regge ansatz, which leads to cancellation of all energy variables
in eq. (\ref{A one-loop}) and finally gives us eq.(\ref{R+L fin}).

In all the above formulae we used the unrenormalized coupling constant $g$.
Threfore, expressions (\ref{R-L fin}) and (\ref{R+L fin}) for $D\rightarrow4$
contain singularities from ultraviolet as well as from infrared divergences of
Feynamn integrals. We can remove the ultraviolet divergences from
$\mathcal{R}-\mathcal{L}$ and $\mathcal{R}+\mathcal{L}$ expressing $g$ in
terms of the renormalized coupling constant, for example in the $\overline
{\mathrm{MS}}$ renormalization scheme%
\begin{equation}
g=g_{\mu}\mu^{2-D/2}\left\{  1-\frac{1}{2}\left(  \frac{11}{3}N-\frac{2}%
{3}n_{f}\right)  \frac{g_{\mu}^{2}\Gamma(2-D/2)}{{(4\pi)}^{D/2}}%
+\cdots\right\}  \,,
\end{equation}
where $n_{f}$ is the number of flavors and $g_{\mu}$ is the renormalized
coupling constant at the renormalization point $\mu$. At first sight, the
absence of the singularity proportional to $n_{f}$ at $D=4$ in our answer for
$\mathcal{R}+\mathcal{L}$ may look suspicious, but one should realize that
such a term with the ultraviolet singularity cancels the infrared singularity
which arises \cite{'t Hooft-72} in the fermion part of gluon self--energy
$\mathcal{P}^{f}\mathcal{(}0\mathcal{)}$ at $m=0$.
%If the
%infrared singularity in (\ref{Pm}) is regularized by mass $m$, then in the
%regularized expression $\left(  \mathcal{R}_{R}^{(1)}+\mathcal{L}_{R}%
%^{(1)}\right)  \mu^{D/2-2}|_{D\rightarrow4}$ we would have $\ln(\frac{\mu}%
%{m})$ instead of $\frac{1}{D-4}-\frac{1}{2}\left(  \ln(4\pi)-\gamma\right)  $.

\begin{ack}
We would like to thank V.S. Fadin for helpful comments and discussions.
\end{ack}

\appendix

\section{Appendix\label{appendix A}}

Here we calculate the transverse momentum tensor integral appearing in
(\ref{R-L integral}) for $D\rightarrow4$. Expressions for its vector
($I_{\Delta}^{\mu(D-2)}$) and scalar ($I_{\Delta}^{(D-2)}$) variants one can
find in Appendix B of \cite{Fadin:1993wh}.%
\begin{align}
I_{\Delta}^{\mu\nu(D-2)}  &  ={\displaystyle\int}\frac{r^{\mu}r^{\nu
}\,\mathtt{d}^{D-2}r_{\bot}}{r_{\bot}^{2}(r+q_{1})_{\bot}^{2}(r+q_{2})_{\bot
}^{2}}\nonumber\\
&  =\Gamma(1+\epsilon)\pi^{D/2-1}\left\{  \frac{g_{\bot}^{\mu\nu}}{2}\left(
\frac{(\mathbf{q}_{1}\mathbf{k})\ln(\mathbf{q}_{2}^{2})}{\mathbf{k}^{2}t_{1}%
}-\frac{(\mathbf{q}_{2}\mathbf{k})\ln(\mathbf{q}_{1}^{2})}{\mathbf{k}^{2}%
t_{2}}-\frac{(\mathbf{q}_{1}\mathbf{q}_{2})\ln(\mathbf{k}^{2})}{t_{1}t_{2}%
}\right)  \right. \nonumber\\
&  -\frac{q_{1\bot}^{\mu}q_{1\bot}^{\nu}}{\mathbf{k}^{2}t_{1}}\left(  \frac
{1}{\epsilon}-\ln(\mathbf{q}_{1}^{2}\mathbf{k}^{2})\right)  -\frac{q_{2\bot
}^{\mu}q_{2\bot}^{\nu}}{\mathbf{k}^{2}t_{2}}\left(  \frac{1}{\epsilon}%
-\ln(\mathbf{q}_{2}^{2}\mathbf{k}^{2})\right) \label{I_3}\\
&  -\left.  \frac{(q_{1}^{\mu}q_{2}^{\nu}+q_{2}^{\mu}q_{1}^{\nu})_{\bot}}%
{2}\left(  \frac{\ln(\mathbf{k}^{2})}{t_{1}t_{2}}+\frac{\ln(\mathbf{q}_{1}%
^{2})}{\mathbf{k}^{2}t_{2}}+\frac{\ln(\mathbf{q}_{2}^{2})}{\mathbf{k}^{2}%
t_{1}}\right)  +\mathcal{O}(\epsilon)\,\right\}  ,\nonumber
\end{align}
Let us explain how one can obtain this short formula. The method used is well
known, and we describe it shortly here to refer to it in the next Section.
First, we introduce our notation for coefficients in the tensor decomposition
of the $I_{\Delta}^{\mu\nu(D-2)}$:
\begin{equation}
I_{\Delta}^{\mu\nu(D-2)}=I_{\Delta}^{(D-2)}[g]\,g_{\bot\lbrack D-2]}^{\mu\nu
}+\sum_{i,j=1}^{2}I_{\Delta}^{(D-2)}[q_{i}q_{j}]\,q_{i\bot}^{\mu}q_{j\bot
}^{\nu}\,, \label{I_3 tensor deconp}%
\end{equation}
where $g_{\bot\lbrack D-2]}^{\mu\nu}$ is the metric tensor in $D-2$
dimensions. These coefficients can be expressed through the vector $I_{\Delta
}^{\mu(D-2)}$ and scalar $I_{\Delta}^{(D)}$integrals:%
\begin{align}
I_{\Delta}^{(D-2)}[q_{1}q_{2}]  &  =\frac{1}{t_{1}t_{2}-(q_{1}q_{2})_{\bot
}^{2}}\left\{  \left[  I_{\Delta}^{(D-2)}[g]-P_{1}^{\mu}\left(  I_{\Delta
}^{\mu\nu(D-2)}q_{1\bot}^{\nu}\right)  \right]  (q_{1}q_{2})_{\bot}+t_{1}%
P_{1}^{\mu}\left(  I_{\Delta}^{\mu\nu(D-2)}q_{2\bot}^{\nu}\right)  \right\}
\,,\nonumber\\
I_{\Delta}^{(D-2)}[q_{1}q_{1}]  &  =\frac{1}{(q_{1}q_{2})_{\bot}^{2}%
-t_{1}t_{2}}\left\{  \left[  I_{\Delta}^{(D-2)}[g]-P_{1}^{\mu}\left(
I_{\Delta}^{\mu\nu(D-2)}q_{1\bot}^{\nu}\right)  \right]  t_{2}+(q_{1}%
q_{2})_{\bot}P_{1}^{\mu}\left(  I_{\Delta}^{\mu\nu(D-2)}q_{2\bot}^{\nu
}\right)  \right\}  \,,\label{I_3 coefficients}\\
I_{\Delta}^{(D-2)}[q_{2}q_{2}]  &  =I_{\Delta}^{(D-2)}[q_{1}q_{1}%
]\,(q_{1}\leftrightarrow q_{2})\,\,,\quad I_{\Delta}^{(D-2)}[g]=\frac{1}{2\pi
}I_{\Delta}^{(D)}\,,\nonumber
\end{align}
where $P_{1}^{\mu}$ is the operator returning the coefficient of $q_{1\bot
}^{\mu}$ in the vector integral $I_{\Delta}^{\mu\nu(D-2)}q_{i\bot}^{\nu},$
i.e.
\begin{equation}
P_{1}^{\mu}=\frac{q_{1\bot}^{\mu}q_{2\bot}^{2}-q_{2\bot}^{\mu}(q_{1}%
q_{2})_{\bot}}{q_{1\bot}^{2}\,q_{2\bot}^{2}-(q_{1}q_{2})_{\bot}^{2}}.
\end{equation}
To find (\ref{I_3 coefficients}) we multiply (\ref{I_3 tensor deconp}) by
$q_{i\bot}^{\nu}$ and solve the equations obtained w.r.t. $I_{\Delta}%
^{(D-2)}[q_{i}q_{j}]$. As one can see, relations (\ref{I_3 coefficients})
contain unpleasant $I_{\Delta}^{(D-2)}[g]=I_{\Delta}^{(D)}/(2\pi)$ expressed
in radicals and polylogarithms \cite{'tHooft:1978xw}. But we notice that
$I_{\Delta}^{(D-2)}[g]$ is finite for $D\rightarrow4$. Therefore, in the
decomposition of the $D-2$ dimensional metric tensor $g_{\bot\lbrack
D-2]}^{\mu\nu}$ in eq.(\ref{I_3 tensor deconp}) into the two--dimensional
metric tensor $g_{\bot\lbrack2]}^{\mu\nu}$ and the metric tensor in $D-4$
dimensions%
\begin{equation}
g_{\bot\lbrack D-2]}^{\mu\nu}=g_{\bot\lbrack2]}^{\mu\nu}+g_{\bot\lbrack
D-4]}^{\mu\nu},
\end{equation}
the last item can be neglected because it gives $\mathcal{O}(\epsilon
)\,$\ contribution. Then, the relation%
\begin{equation}
((q_{1}q_{2})_{\bot}^{2}-t_{1}t_{2})g_{\bot\lbrack2]}^{\mu\nu}=(q_{1}^{\mu
}q_{2}^{\nu}+q_{2}^{\mu}q_{1}^{\nu})_{\bot}(q_{1}q_{2})_{\bot}-q_{1\bot}^{\mu
}q_{1\bot}^{\nu}t_{2}-q_{2\bot}^{\mu}q_{2\bot}^{\nu}t_{1} \label{g decompose}%
\end{equation}
helps us totally cancel $I_{\Delta}^{(D-2)}[g]$ in (\ref{I_3 tensor deconp}).
Finally, we eliminate all Gram determinants $(q_{1,}q_{2})_{\bot}^{2}%
-t_{1}t_{2}$ from denominators of $I_{\Delta}^{(D-2)}[q_{i}q_{j}]$ by the
appropriate choice of a new $I_{\Delta}^{(D-2)}[g]$ , see (\ref{I_3}).

\section{Appendix\label{appendix B}}

Here we calculate in the Regge limit some of the integrals appearing in
eqs.(\ref{A2-3(as)}) and (\ref{symmetr add}). Other necessary integrals can be
found in Appendix C of \cite{Fadin:1993wh} and Appendices A and B of
\cite{Fadin:2000yp}. We use the following denotations:%
\begin{equation}
\quad c_{\Gamma}=\frac{\Gamma(1+\epsilon)\Gamma(1-\epsilon)^{2}}%
{\Gamma(1-2\epsilon)(4\pi)^{D/2}}\,,\quad(-k_{12})\equiv\frac{(-s_{1}%
)(-s_{2})}{(-s)}\,,\quad I(t)=\frac{c_{\Gamma}(-t)^{-\epsilon}}{\epsilon
(1-2\epsilon)}\,, \label{c-Gamma}%
\end{equation}%
\begin{align}
I_{4}^{\mu}  &  =%
%TCIMACRO{\dint }%
%BeginExpansion
{\displaystyle\int}
%EndExpansion
\frac{\,\mathtt{d}^{D}r}{\mathtt{i}(2\pi)^{D}}\frac{r^{\mu}}{r^{2}p^{2}%
s_{A}s_{B}}\nonumber\\
&  =p_{A}^{\mu}\frac{2(D-3)}{(D-4)}\frac{I(t_{1})}{st_{1}}+p_{B}^{\mu}%
\frac{2(D-3)}{(D-4)}\frac{I(t_{1})-I(s_{2})}{ss_{2}}-q_{1}^{\mu}\left(
\frac{2(D-3)}{(D-4)}\frac{I(t_{1})}{st_{1}}+\frac{I_{4}}{2}\right)  \,,
\label{I_4 mu}%
\end{align}
where
\begin{equation}
I_{4}=%
%TCIMACRO{\dint }%
%BeginExpansion
{\displaystyle\int}
%EndExpansion
\frac{\,\mathtt{d}^{D}r}{\mathtt{i}(2\pi)^{D}}\frac{1}{r^{2}p^{2}s_{A}s_{B}%
}=\frac{c_{\Gamma}}{\epsilon}\frac{2}{s(-t_{1})^{1+\epsilon}}\left(
\ln\left(  \frac{-s}{-s_{2}}\right)  +\psi(1)-\psi(-\epsilon)\right)  \,
\end{equation}
may be found in \cite{Fadin:2000yp} and $\psi(z)=\Gamma^{\prime}(z)/\Gamma
(z)$.%
\begin{equation}
I_{4A}^{\mu}=%
%TCIMACRO{\dint }%
%BeginExpansion
{\displaystyle\int}
%EndExpansion
\frac{\,\mathtt{d}^{D}r}{\mathtt{i}(2\pi)^{D}}\frac{r^{\mu}}{r^{2}p^{2}ns_{A}%
}=p_{A}^{\mu}\frac{2(D-3)}{(D-4)}\frac{I(t_{1})}{s_{1}t_{1}}-k^{\mu}%
\frac{2(D-3)}{(D-4)s_{1}}\frac{I(t_{1})-I(t_{2})}{t_{1}-t_{2}}-q_{1}^{\mu
}\frac{1}{2}I_{4\text{A}}\,,
\end{equation}%
\begin{align}
I_{4A}^{\mu\nu}  &  =%
%TCIMACRO{\dint }%
%BeginExpansion
{\displaystyle\int}
%EndExpansion
\frac{\,\mathtt{d}^{D}r}{\mathtt{i}(2\pi)^{D}}\frac{r^{\mu}r^{\nu}}{r^{2}%
p^{2}ns_{A}}\nonumber\\
&  =(k^{\mu}q_{1}^{\nu}+k^{\nu}q_{1}^{\mu})\frac{(D-2)}{(D-4)s_{1}}%
\frac{I(t_{1})-I(t_{2})}{t_{1}-t_{2}}+q_{1}^{\mu}q_{1}^{\nu}\frac{(D-2)I_{4A}%
}{4(D-3)}\nonumber\\
&  +\frac{k^{\mu}k^{\nu}}{s_{1}}\left(  \frac{I(t_{2})}{t_{1}-t_{2}}%
-\frac{2t_{1}(I(t_{1})-I(t_{2}))}{(D-4)\left(  t_{1}-t_{2}\right)  ^{2}%
}\right)  -p_{A}^{\mu}p_{A}^{\nu}\frac{2I(t_{1})}{(D-4)s_{1}t_{1}%
}\label{I_4Amu nu}\\
&  +g^{\mu\nu}\left(  \frac{I(t_{1})-I(t_{2})+I(s_{1})}{(4-D)s_{1}}%
+\frac{t_{1}I_{4A}}{4(3-D)}\right)  -(p_{A}^{\mu}q_{1}^{\nu}+p_{A}^{\nu}%
q_{1}^{\mu})\frac{I(t_{1})}{s_{1}t_{1}}\nonumber\\
&  +(p_{A}^{\nu}k^{\mu}+p_{A}^{\mu}k^{\nu})\left(  \frac{(D-2)I(t_{1}%
)-2I(t_{2})}{(D-4)s_{1}^{2}}+\frac{t_{1}I_{4A}}{2(D-3)s_{1}}-\frac
{(D-6)I(s_{1})}{(D-4)s_{1}^{2}}\right)  \,,\nonumber
\end{align}
where $I_{4A}$ may be found in \cite{Fadin:1993wh}%
\begin{align}
I_{4A}  &  =%
%TCIMACRO{\dint }%
%BeginExpansion
{\displaystyle\int}
%EndExpansion
\frac{\,\mathtt{d}^{D}r}{\mathtt{i}(2\pi)^{D}}\frac{1}{r^{2}p^{2}ns_{A}%
}\nonumber\\
&  =\frac{c_{\Gamma}}{s_{1}t_{1}}\frac{\Gamma(1-2\epsilon)}{\Gamma
(1-\epsilon)^{2}}\left(  \frac{2}{\epsilon^{2}}-\frac{2}{\epsilon}\ln\left(
\frac{(-s_{1})(-t_{1})}{(-t_{2})}\right)  \right. \\
&  -\left.  \ln^{2}\left(  -t_{2}\right)  +2\ln\left(  -s_{1}\right)
\ln\left(  -t_{1}\right)  -2\Li\left(  1-\frac{t_{2}}{t_{1}}\right)  -\pi
^{2}\right)  +\mathcal{O}(\epsilon)\,.\nonumber
\end{align}

The most complicated of appearing integrals is a tensor pentabox, which we
present in $D\rightarrow4$ decomposition:%
\begin{align*}
&  c_{\Gamma}^{-1}I_{5}^{\mu\nu}=c_{\Gamma}^{-1}%
%TCIMACRO{\dint }%
%BeginExpansion
{\displaystyle\int}
%EndExpansion
\frac{\,\mathtt{d}^{D}r}{\mathtt{i}(2\pi)^{D}}\frac{r^{\mu}r^{\nu}}{r^{2}%
p^{2}ns_{A}s_{B}}\\
&  =\frac{g^{\mu\nu}}{8\,s_{1}s_{2}}\left\{  \frac{\left(  t_{1}%
+t_{2}+\mathbf{k}^{2}\right)  ^{2}-4t_{1}t_{2}}{t_{1}t_{2}}\ln^{2}\left(
\frac{-k_{12}}{\mathbf{k}^{2}}\right)  +\frac{t_{2}-t_{1}-\mathbf{k}^{2}%
}{t_{2}}\ln^{2}\left(  \frac{-k_{12}}{-t_{1}}\right)  \right. \\
&  +\frac{t_{1}-t_{2}-\mathbf{k}^{2}}{t_{1}}\left(  \ln^{2}\left(
\frac{-k_{12}}{-t_{2}}\right)  -\ln^{2}\left(  \frac{-t_{1}}{-t_{2}}\right)
\right)  -\frac{\left(  t_{1}-t_{2}\right)  \left(  t_{1}+t_{2}+\mathbf{k}%
^{2}\right)  }{t_{1}t_{2}}2\Li\left(  1-\frac{t_{2}}{t_{1}}\right) \\
&  +\left.  \frac{\pi^{2}}{3}\frac{(\mathbf{k}^{2})^{2}-\left(  t_{1}%
-t_{2}\right)  ^{2}}{t_{1}t_{2}}\right\}
\end{align*}
\vspace{-0.8cm}
\begin{align*}
&  +\,\frac{p_{A}^{\mu}p_{B}^{\nu}+p_{B}^{\mu}p_{A}^{\nu}}{4\,s\,s_{1}\,s_{2}%
}\left\{  -\frac{\left(  t_{1}-t_{2}\right)  ^{2}+\mathbf{k}^{2}\left(
t_{1}+t_{2}\right)  }{t_{1}t_{2}}\ln^{2}\left(  \frac{-k_{12}}{\mathbf{k}^{2}%
}\right)  \right. \\
&  +\left(  t_{1}-t_{2}\right)  \left(  \frac{1}{t_{2}}\ln^{2}\left(
\frac{-k_{12}}{-t_{1}}\right)  -\frac{1}{t_{1}}\ln^{2}\left(  \frac{-k_{12}%
}{-t_{2}}\right)  +\frac{1}{t_{1}}\ln^{2}\left(  \frac{-t_{1}}{-t_{2}}\right)
\right)  +\\
&  +\left.  \frac{t_{1}^{2}-t_{2}^{2}}{t_{1}t_{2}}2\Li\left(  1-\frac{t_{2}%
}{t_{1}}\right)  +\frac{\pi^{2}}{3}\frac{\left(  t_{1}-t_{2}\right)
^{2}-\mathbf{k}^{2}\left(  t_{1}+t_{2}\right)  }{t_{1}t_{2}}\right\}
\end{align*}
\vspace{-0.8cm}
\begin{align*}
&  +\frac{q_{1}^{\mu}q_{2}^{\nu}+q_{2}^{\mu}q_{1}^{\nu}}{4\,s_{1}\,s_{2}%
}\left\{  \frac{-4}{\epsilon\left(  t_{1}-t_{2}\right)  }\ln\left(
\frac{-t_{1}}{-t_{2}}\right)  -\frac{1}{t_{1}}\ln^{2}\left(  \frac{-t_{1}%
}{-t_{2}}\right)  \right. \\
&  +\frac{2}{t_{1}-t_{2}}\left(  \ln^{2}\left(  -t_{1}\right)  -\ln^{2}\left(
-t_{2}\right)  \right)  +\frac{t_{1}-t_{2}}{t_{1}t_{2}}2\Li\left(
1-\frac{t_{2}}{t_{1}}\right)  +\frac{\pi^{2}}{3}\frac{t_{1}+t_{2}%
-\mathbf{k}^{2}}{t_{1}t_{2}}\\
&  +\left.  \frac{1}{t_{2}}\ln^{2}\left(  \frac{-k_{12}}{-t_{1}}\right)
+\frac{1}{t_{1}}\ln^{2}\left(  \frac{-k_{12}}{-t_{2}}\right)  -\frac
{\mathbf{k}^{2}+t_{1}+t_{2}}{t_{1}t_{2}}\ln^{2}\left(  \frac{-k_{12}%
}{\mathbf{k}^{2}}\right)  \right\}
\end{align*}
\vspace{-0.85cm}
\begin{align*}
&  +\left[  \frac{p_{A}^{\mu}q_{1}^{\nu}+p_{A}^{\nu}q_{1}^{\mu}}{s\,s_{1}%
t_{1}}\left\{  \frac{1}{\epsilon}\ln\left(  \frac{-k_{12}}{-s_{1}}\right)
-\frac{1}{2}\ln^{2}(-k_{12})-\frac{1}{2}\ln^{2}\left(  \frac{-k_{12}%
}{\mathbf{k}^{2}}\right)  +\frac{1}{2}\ln^{2}\left(  \frac{-k_{12}}{-t_{1}%
}\right)  \right.  \right. \\
&  +\left.  \frac{1}{2}\ln^{2}\left(  -s_{1}\right)  -\frac{1}{2}\ln
^{2}\left(  \frac{-s_{1}}{-t_{1}}\right)  -\frac{\pi^{2}}{3}\right\}
+\,\frac{p_{A}^{\mu}p_{A}^{\nu}}{s\,s_{1}\,t_{1}}\left\{  \frac{1}%
{\epsilon^{2}}-\frac{1}{\epsilon}\ln\left(  -t_{1}\right)  +\frac{1}{2}\ln
^{2}\left(  -t_{1}\right)  \right\}
\end{align*}
\vspace{-0.8cm}
\begin{align*}
&  +\,\frac{p_{A}^{\mu}q_{2}^{\nu}+p_{A}^{\nu}q_{2}^{\mu}}{4\,s\,s_{1}%
}\left\{  \frac{t_{1}+t_{2}+\mathbf{k}^{2}}{t_{1}t_{2}}\ln^{2}\left(
\frac{-k_{12}}{\mathbf{k}^{2}}\right)  +\frac{1}{t_{1}}\ln^{2}\left(
\frac{-t_{1}}{-t_{2}}\right)  -\frac{1}{t_{1}}\ln^{2}\left(  \frac{-k_{12}%
}{-t_{2}}\right)  \right. \\
&  -\left.  \frac{1}{t_{2}}\ln^{2}\left(  \frac{-k_{12}}{-t_{1}}\right)
-\frac{t_{1}-t_{2}}{t_{1}t_{2}}2\Li\left(  1-\frac{t_{2}}{t_{1}}\right)
-\frac{\pi^{2}}{3}\frac{t_{1}+t_{2}-\mathbf{k}^{2}}{t_{1}t_{2}}\right\}
\end{align*}
\vspace{-0.8cm}
\begin{align}
&  +\frac{q_{1}^{\mu}q_{1}^{\nu}}{s_{1}\,s_{2}\,t_{1}}\left\{  \frac
{1}{\epsilon^{2}}+\frac{1}{\epsilon}\left[  \frac{t_{2}}{t_{1}-t_{2}}%
\ln\left(  \frac{-t_{1}}{-t_{2}}\right)  -\ln(-k_{12})\right]  -\frac{1}{2}%
\ln^{2}\left(  \frac{-k_{12}}{-t_{1}}\right)  \right. \label{I_5 mu nu}\\
&  +\frac{1}{2}\ln^{2}(-k_{12})+\frac{1}{2}\ln^{2}\left(  \frac{-k_{12}%
}{\mathbf{k}^{2}}\right)  -\frac{t_{2}}{2\left(  t_{1}-t_{2}\right)  }\left(
\ln^{2}\left(  -t_{1}\right)  -\ln^{2}\left(  -t_{2}\right)  \right)
\nonumber\\
&  -\left.  \left.  \Li\left(  1-\frac{t_{2}}{t_{1}}\right)  \right\}
+(1\leftrightarrow2)\right]  +\mathcal{O}(\epsilon)\,,\nonumber
\end{align}
where $(1\leftrightarrow2)$ means $p_{A}\leftrightarrow-p_{B}$, and
$q_{1}\leftrightarrow q_{2}$ substitutions.

To find (\ref{I_5 mu nu}) we perform a procedure similar to the one used for
calculating $I_{\Delta}^{\mu\nu(D-2)}$ in Appendix \ref{appendix A}. Again,
the coefficient of metric tensor is expressed through scalar pentabox in $D+2$
dimensions and, therefore, is finite, which allows one totally cancel it using
the relations analogous to (\ref{I_3 coefficients}) and (\ref{g decompose}).
This cancellation looks much simpler in Sudakov basis: $p_{A}$, $p_{B}$,
$q_{i\bot}$, there corresponding formulae are very similar to
eqs.(\ref{I_3 coefficients}) and%
\begin{equation}
g_{[4]}^{\mu\nu}=\frac{p_{A}^{\mu}p_{B}^{\nu}+p_{A}^{\nu}p_{B}^{\mu}}%
{(p_{A}p_{B})}+g_{\bot\lbrack2]}^{\mu\nu}\,, \label{g[4] decomp}%
\end{equation}
with $g_{\bot\lbrack2]}^{\mu\nu}$ from (\ref{g decompose}). Then, most of
remaining vector integrals are known \cite{Fadin:1993wh,Fadin:2000yp}, and the
others may be calculated by Sudakov decomposition technique exploited in
\cite{Fadin:2000yp}. Fortunately, the analysis of $I_{5}^{\mu\nu}$ in Feynman
parametrization shows that it does not contain terms proportional to $s_{i}$
in fractional power, such as $I(s_{1})$ in $I_{4A}^{\mu\nu}$. That is why our
answer coincides with the one obtained in \cite{Bern:1993kr} where the limit
$D\rightarrow4$ was taken first. Then, we use freedom in choice of the
coefficient of $g^{\mu\nu}$ and choose it so (see eq.(\ref{I_5 mu nu})) that
on the one side it eliminates all Gram determinants $(q_{1}q_{2})_{\bot}%
^{2}-t_{1}t_{2}$ in denominators, and on the other side, contains energy
dependence compatible with Reggeization after its insertion in
(\ref{A one-loop}). Moreover, in calculating (\ref{R-L answer 1}) the
analogous freedom was exploited (see Appendix \ref{appendix A}). So, to
perform the necessary cancellations of terms proportional to $\mathcal{R}%
^{(0)}-\mathcal{L}^{(0)}$ in (\ref{A one-loop}) we have to choose in
(\ref{I_5 mu nu}) the coefficient of $g^{\mu\nu}$ in a way consistent with
corresponding coefficient of $I_{\Delta}^{\mu\nu(D-2)}$. Because of this
freedom, formula (\ref{I_5 mu nu}) does not straightforwardly coincide with
the corresponding one presented in \cite{Campbell:1996zw}, there $\ln
^{2}(s_{i})$ may be found. Such $\ln^{2}(s_{i})$ are in contradiction with
Regge asymptotic (\ref{A one-loop}) but may be eliminated via discussed trick.
At last, we present one additional vector integral useful for the calculation
of $I_{5}^{\mu\nu}$%
\begin{align}
I_{4k}^{\mu}  &  =%
%TCIMACRO{\dint }%
%BeginExpansion
{\displaystyle\int}
%EndExpansion
\frac{\,\mathtt{d}^{D}r}{\mathtt{i}(2\pi)^{D}}\frac{r^{\mu}}{p^{2}ns_{A}s_{B}%
}\nonumber\\
&  =p_{A}^{\mu}\left(  \frac{s_{2}}{2s}I_{4k}+\frac{2(D-3)}{(D-4)ss_{1}%
}I(s_{1})\right)  -p_{B}^{\mu}\left(  \frac{s_{1}}{2s}I_{4k}+\frac
{2(D-3)}{(D-4)ss_{2}}I(s_{2})\right)  -\frac{1}{2}\left(  q_{1}^{\mu}%
+q_{2}^{\mu}\right)  I_{4k}\,,
\end{align}
where its scalar variant $I_{4k}$ also appears in (\ref{symmetr add}) and may
be found in \cite{Fadin:2000yp}%
\[
I_{4k}=%
%TCIMACRO{\dint }%
%BeginExpansion
{\displaystyle\int}
%EndExpansion
\frac{\,\mathtt{d}^{D}r}{\mathtt{i}(2\pi)^{D}}\frac{1}{p^{2}ns_{A}s_{B}}%
=\frac{c_{\Gamma}}{s(\mathbf{k}^{2})^{1+\epsilon}}\frac{2}{\epsilon}\left[
\ln\left(  \frac{(-s)\mathbf{k}^{2}}{\mathbf{(-}s_{1})(-s_{2})}\right)
+\psi(-\epsilon)-\psi(1+\epsilon)\right]  \,.
\]

\section{Appendix\label{appendix C}}

Here we present vector and tensor integrals necessary to calculate
$\Gamma_{A^{\prime}A}^{(1)}$ (see (\ref{Gamma integrand})):
\begin{align*}
I_{\square}^{\mu\nu}  &  =%
%TCIMACRO{\dint }%
%BeginExpansion
{\displaystyle\int}
%EndExpansion
\frac{\,\mathtt{d}^{D}r}{\mathtt{i}(2\pi)^{D}}\frac{r^{\mu}r^{\nu}}{r^{2}%
p^{2}s_{A}s_{B}}=-(p_{A}^{\mu}p_{A}^{\nu}+p_{B}^{\mu}p_{B}^{\nu})\frac
{2I(q)}{(D-4)st}-g^{\mu\nu}\left(  \frac{I(t)+I(s)}{(D-4)s}+\frac
{t\,I_{\square}}{4(D-3)}\right) \\
&  +q^{\mu}q^{\nu}\left(  \frac{2(D-3)I(t)}{(D-4)st}+\frac{(D-2)I_{\square}%
}{4(D-3)}\right)  -((p_{A}-p_{B})^{\mu}q^{\nu}+(p_{A}-p_{B})^{\nu}q^{\mu
})\frac{I(q)}{s\,t}\\
&  +\frac{p_{A}^{\mu}p_{B}^{\nu}+p_{A}^{\nu}p_{B}^{\mu}}{s}\left(
\frac{(D-2)I(t)-(D-6)I(s)}{(D-4)s}+\frac{t\,I_{\square}}{2(D-3)}\right)  \,,
\end{align*}%
\begin{equation}
I_{\square}^{\mu}=%
%TCIMACRO{\dint }%
%BeginExpansion
{\displaystyle\int}
%EndExpansion
\frac{\,\mathtt{d}^{D}r}{\mathtt{i}(2\pi)^{D}}\frac{r^{\mu}}{r^{2}p^{2}%
s_{A}s_{B}}=-\frac{q^{\mu}}{2}I_{\square}+\frac{2(D-3)}{D-4}\frac
{(p_{A}-q-p_{B})^{\mu}}{s\,t}\,I(t),
\end{equation}
where
\begin{equation}
I_{\square}=%
%TCIMACRO{\dint }%
%BeginExpansion
{\displaystyle\int}
%EndExpansion
\frac{\,\mathtt{d}^{D}r}{\mathtt{i}(2\pi)^{D}}\frac{1}{r^{2}(r+q)^{2}%
s_{A}s_{B}}=\frac{2}{s(-t)^{1+\epsilon}}\frac{c_{\Gamma}}{\epsilon}\left\{
\ln\left(  \frac{-s}{-t}\right)  -2\psi(-\epsilon)+\psi(1+\epsilon
)+\psi(1)\right\}  \,,
\end{equation}
was calculated in \cite{Fadin:1993wh}, and for $c_{\Gamma}$ see
eq.(\ref{c-Gamma}).

\end{document}